\documentclass[english,aps,prd,nofootinbib,showkeys,preprint,floatfix]{revtex4}

\usepackage{amsmath}
\usepackage{amssymb}
\usepackage{bbm}
\usepackage{graphicx}
\usepackage{rotating}
\usepackage{color} 
\usepackage{multirow} 
\usepackage[T1]{fontenc} 
\usepackage[latin9]{inputenc}
\usepackage{lmodern}
\makeatletter
\providecommand{\tabularnewline}{\\}
\usepackage{booktabs}
\usepackage{mathrsfs}   
\usepackage{slashed}     
\usepackage{url}
\usepackage{multirow}
\usepackage{units}

\usepackage{ wasysym }

\makeatother

\usepackage{babel}

\newcommand {\Tr} {{\rm Tr}}
\newcommand {\be} {\begin{equation}}
\newcommand {\ee} {\end{equation}}

\newcommand{\nn}{\nonumber}

\newcommand{\AddrAHEP}{
  {\it AHEP Group, Instituto de F\'{\i}sica Corpuscular --
    C.S.I.C./Universitat de Val{\`e}ncia \\
    Edificio de Institutos de Paterna, Apartado 22085,
  E--46071 Val{\`e}ncia, Spain}}

\newcommand{\AddrLisb}{%
 Departamento de F\'\i sica and CFTP, Instituto Superior T\'ecnico\\
 Universidade de Lisboa, 
          Av. Rovisco Pais 1, 1049-001 Lisboa, Portugal }

\newcommand{\AddrPraha}{%
Institute of Particle and Nuclear Physics, Faculty of Mathematics and 
Physics,\\ Charles University in Prague, V Hole\v{s}ovi\v{c}k\'ach 2, 
180 00 Praha 8, Czech Republic}

\def\gsim{\raise0.3ex\hbox{$\;>$\kern-0.75em\raise-1.1ex\hbox{$\sim\;$}}}
\def\lsim{\raise0.3ex\hbox{$\;<$\kern-0.75em\raise-1.1ex\hbox{$\sim\;$}}}

\begin{document}

\preprint{CFTP/13-025 and IFIC/13-84}

\title{LHC-scale left-right symmetry and unification}

\author{Carolina Arbel\'aez}\email{carolina@ific.uv.es}
\affiliation{\AddrLisb} 
\affiliation{\AddrAHEP}
\author{Jorge C. Rom\~ao}\email{jorge.romao@tecnico.ulisboa.pt}
\affiliation{\AddrLisb} 
\author{Martin Hirsch} \email{mahirsch@ific.uv.es}
\affiliation{\AddrAHEP}
\author{Michal Malinsk\'y} \email{malinsky@ipnp.troja.mff.cuni.cz}
\affiliation{\AddrPraha}


\begin{abstract}
We construct a comprehensive list of non-supersymmetric standard model
extensions with a low-scale LR-symmetric intermediate stage that may
be obtained as simple low-energy effective theories within a class of
renormalizable $SO(10)$ GUTs.  Unlike the traditional ``minimal'' LR
models many of our example settings support a perfect gauge coupling
unification even if the LR scale is in the LHC domain at a price of
only (a few copies of) one or two types of extra fields pulled down to the
TeV-scale ballpark.  We discuss the main aspects of a potentially
realistic model building conforming the basic constraints from the
quark and lepton sector flavour structure, proton decay limits,
etc. We pay special attention to the theoretical uncertainties
related to the limited information about the underlying unified
framework in the bottom-up approach, in particular, to their role in
the possible extraction of the LR-breaking scale. We observe a general
tendency for the models without new coloured states in the TeV domain
to be on the verge of incompatibility with the proton stability
constraints.

\end{abstract}

\maketitle

\section{Introduction}

It is well-known that with only the standard model (SM) field 
content the gauge couplings do not unify at a single energy scale, 
while the minimal supersymmetric standard model (MSSM) 
leads to quantitatively precise gauge coupling unification 
(GCU), if the scale of supersymmetry is ``close'' to the electro-weak (EW) 
scale \cite{Dimopoulos:1981yj,Ibanez:1981yh,Marciano:1981un,Einhorn:1981sx,Amaldi:1991cn,Langacker:1991an,Ellis:1990wk}.
\footnote{Actually, within supersymmetric models it is only required
that the new fermions (higgsinos, wino and gluino) have masses near
the EW scale, as in the so-called ``split SUSY'' scenario
\cite{ArkaniHamed:2004fb,Giudice:2004tc}.}  However, there are many
extensions of the SM that lead to GCU without supersymmetry (SUSY). In
particular, it is much less known that already in \cite{Amaldi:1991zx}
GCU was studied in a number of non-SUSY extensions of the SM. We also
mention one particular example with vector-like quarks (VLQ) that was
discussed recently in \cite{Gogoladze:2010in}, where the Higgs mass
and stability bounds and the GCU were considered in an SM extension
with two different VLQs.

On the other hand, there are rather few publications which discuss GCU
within left-right symmetric extensions of the SM.  The main reason for
this is probably the fact that for minimal left-right (LR) symmetric
extensions of the SM the couplings do not unify unless the LR scale is
rather high, say ($10^9-10^{11}$) GeV, as has been shown already in
\cite{Brahmachari:1991np}. 

While for the SM the term ``minimal'' is unambiguously defined, for LR
symmetric extensions of the SM the term ``minimal-LR'' model has been
used for quite different models in the literature. Usually in
``minimal LR'' models a second SM Higgs doublet is added to the SM
field content at the LR scale to complete a bi-doublet,
$\Phi_{1,2,2,0}$,
\footnote{Throughout this paper we will use the notation $\Phi$ for
scalars and $\Psi$ for fermions with the subscript denoting the quantum
numbers with respect to either the left-right ($SU(3)_c \times SU(2)_L
\times SU(2)_R \times U(1)_{B-L}$) or the SM group ($SU(3)_c \times
SU(2)_L \times U(1)_{Y}$).}  as required by the LR group. To break the
LR group to the SM group one then (usually) adds a pair of triplets
$\Phi_{1,3,1,-2}$ + $\Phi_{1,1,3,-2}$
\cite{Mohapatra:1977mj,Mohapatra:1979ia,Mohapatra:1980yp}. Here the
presence of the left-triplet $\Phi_{1,3,1,-2}$ allows to maintain
parity in the LR phase, i.e. $g_L=g_R$, sometimes also called
``manifest LR'' symmetry. This construction automatically also creates
a seesaw mass for the right-handed neutrinos from the vacuum
expectation value of the $\Phi_{1,1,3,-2}$ \cite{Mohapatra:1980yp}.
We will call this setup the ``minimal LR'' (mLR) model in the
following.  Alternatively, also a pair of doublets, $\Phi_{1,2,1,-1}$
+ $\Phi_{1,1,2,-1}$, could break the LR group for an equally simple
setup. However, in this case one would need to rely on an inverse 
\cite{Mohapatra:1986bd} (or linear \cite{Akhmedov:1995ip,Akhmedov:1995vm}) 
seesaw for generating neutrino masses. 

In \cite{Brahmachari:2003wv} it has been argued that a ``truly minimal
LR model'' has only two doublets $\Phi_{1,2,1,-1}$ + $\Phi_{1,1,2,-1}$
but no bi-doublet.  In this case, all fermion masses are generated
from non-renormalizable operators (NROs). While this setup has indeed
one field less than the above ``minimal-LR'' models, it needs some
additional unspecified new physics to generate the NROs and, thus, can
not be considered a complete model. Unification in this ``truly
minimal'' setup is achieved for an LR scale around roughly $10^8$ GeV
(and a grand unified theory (GUT) scale of roughly $10^{15}$ GeV
\cite{Siringo:2012bc}. 

\begin{figure}[t]
\centering
\includegraphics[width=0.50\linewidth]{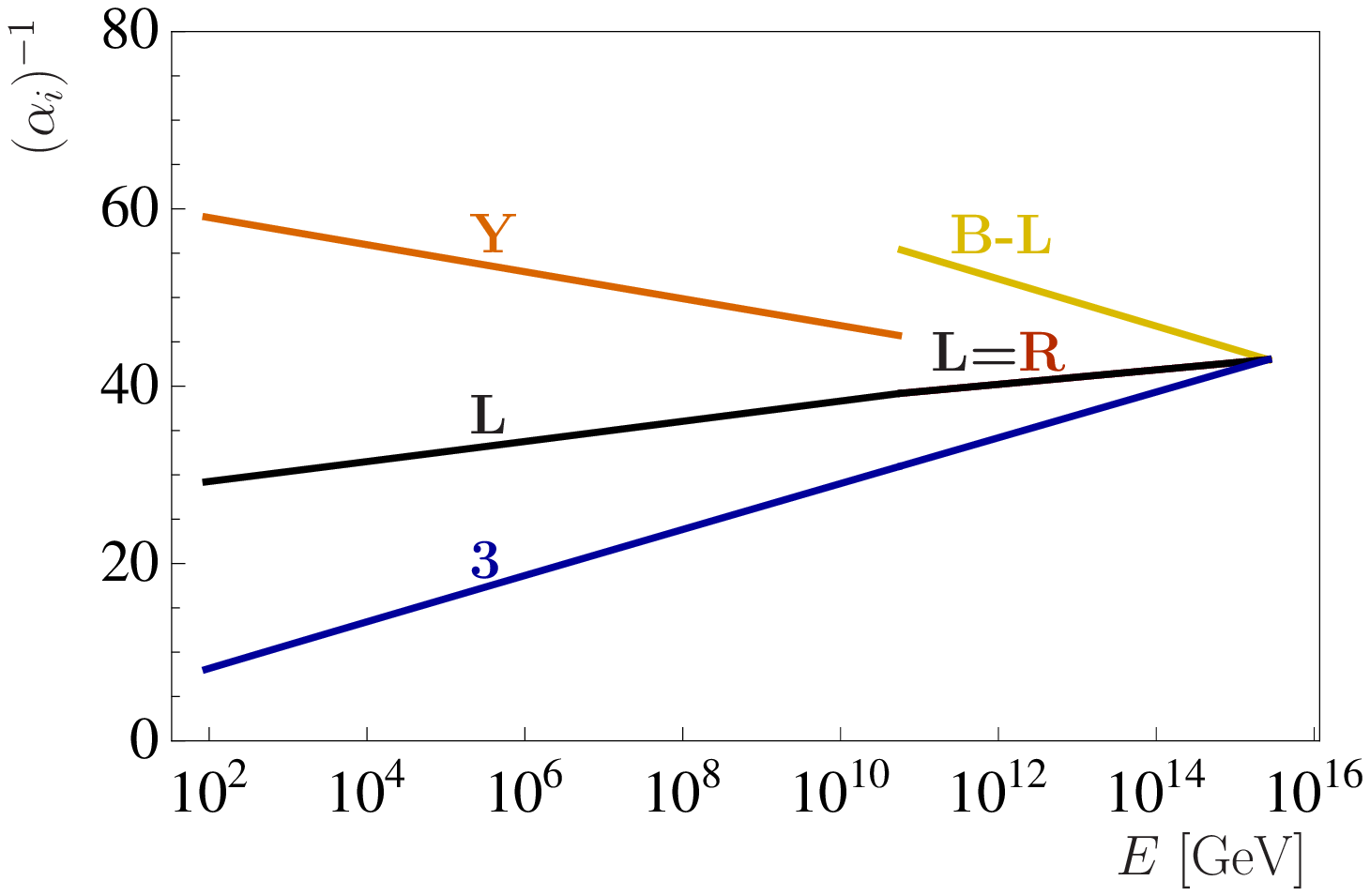}
\hskip-2mm\includegraphics[width=0.50\linewidth]{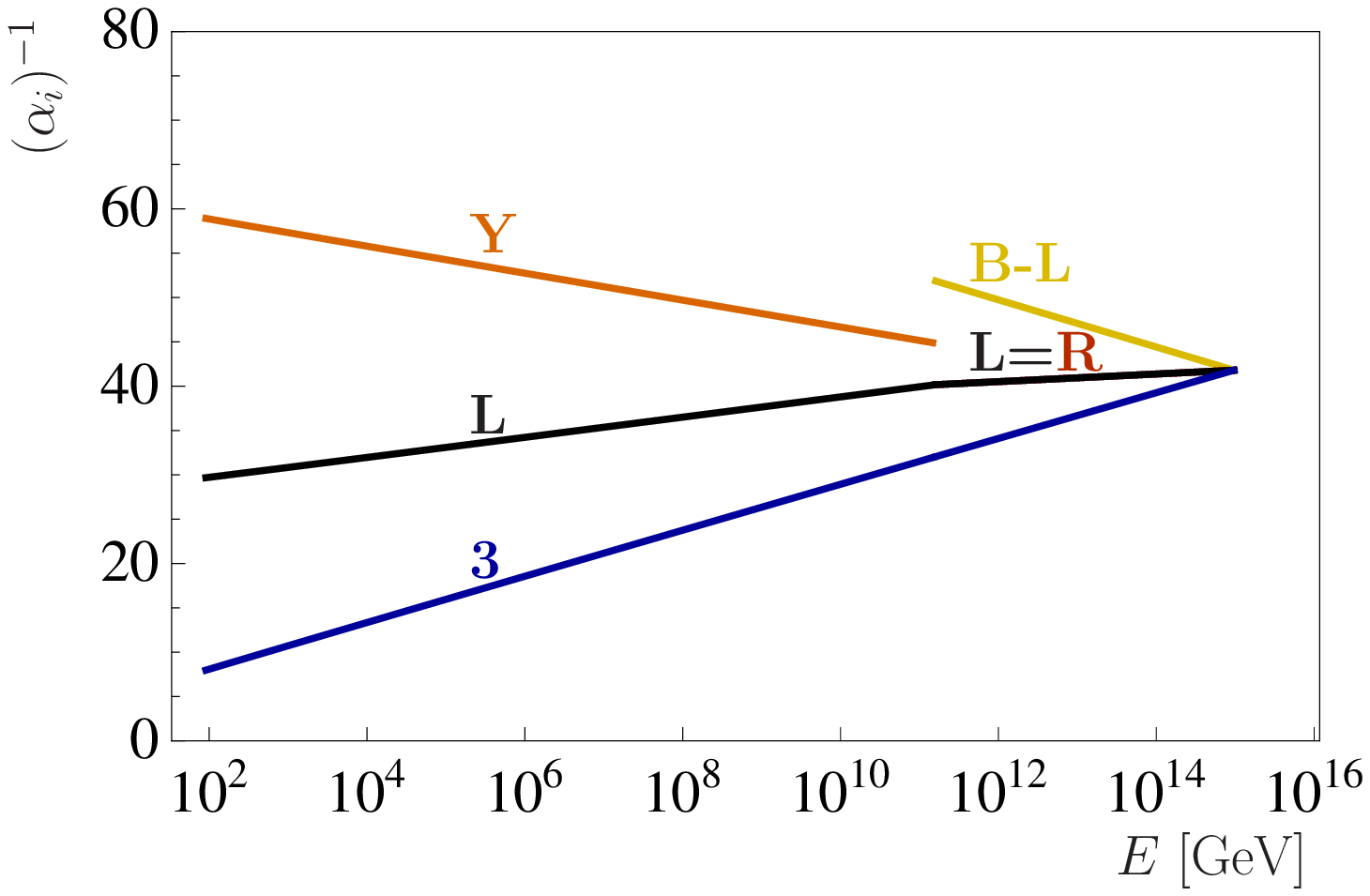}
\caption{Gauge coupling unification, including the 2-loop
$\beta$-coefficients, for two ``minimal'' left-right models, 
to the left ``mLR'', to the right ``m$\Omega$LR'' model. 
For definition of the models and discussion see text.}
\label{fig:MLR}
\end{figure}

A LR model with only bi-doublets can not generate the observed
Cabibbo-Kobayashi-Maskawa (CKM) mixing angles at tree-level, see the
discussion in the next section. This can be solved by adding a second
$\Phi_{1,2,2,0}$ plus a pair of $(B-L)$ neutral triplets,
$\Phi_{1,3,1,0}$ + $\Phi_{1,1,3,0}$.  A supersymmetric version of this
setup has been discussed in \cite{Aulakh:1997ba,Aulakh:1997fq}, see
also \cite{Esteves:2011gk}.  We will call this model the ``minimal
$\Omega$LR'' (m$\Omega$LR) model.  Fig.~\ref{fig:MLR} shows the
running of the gauge couplings for the minimal setup (``mLR''),
including 2-loop beta coefficients, in the left plot and for the
m$\Omega$LR model in the right plot.  Note, that the best fit point
(b.f.p.) for $m_{LR} = 3 \times 10^{10}$ GeV and 
$m_{G} = 2 \times 10^{15}$ GeV in the mLR model, while the b.f.p. for
$m_{LR} = 3 \times 10^{11}$ GeV and $m_{G} = 6 \times 10^{14}$ GeV in
the m$\Omega$LR model.\footnote{The authors of
  \cite{Aulakh:1997ba,Aulakh:1997fq} called this the ``minimal
  supersymmetric LR'' model. In this original supersymmetric version
  the b.f.p. for the LR scale from GCU is equal to the GUT scale.}

\begin{figure}[t]
\centering
\includegraphics[width=0.50\linewidth]{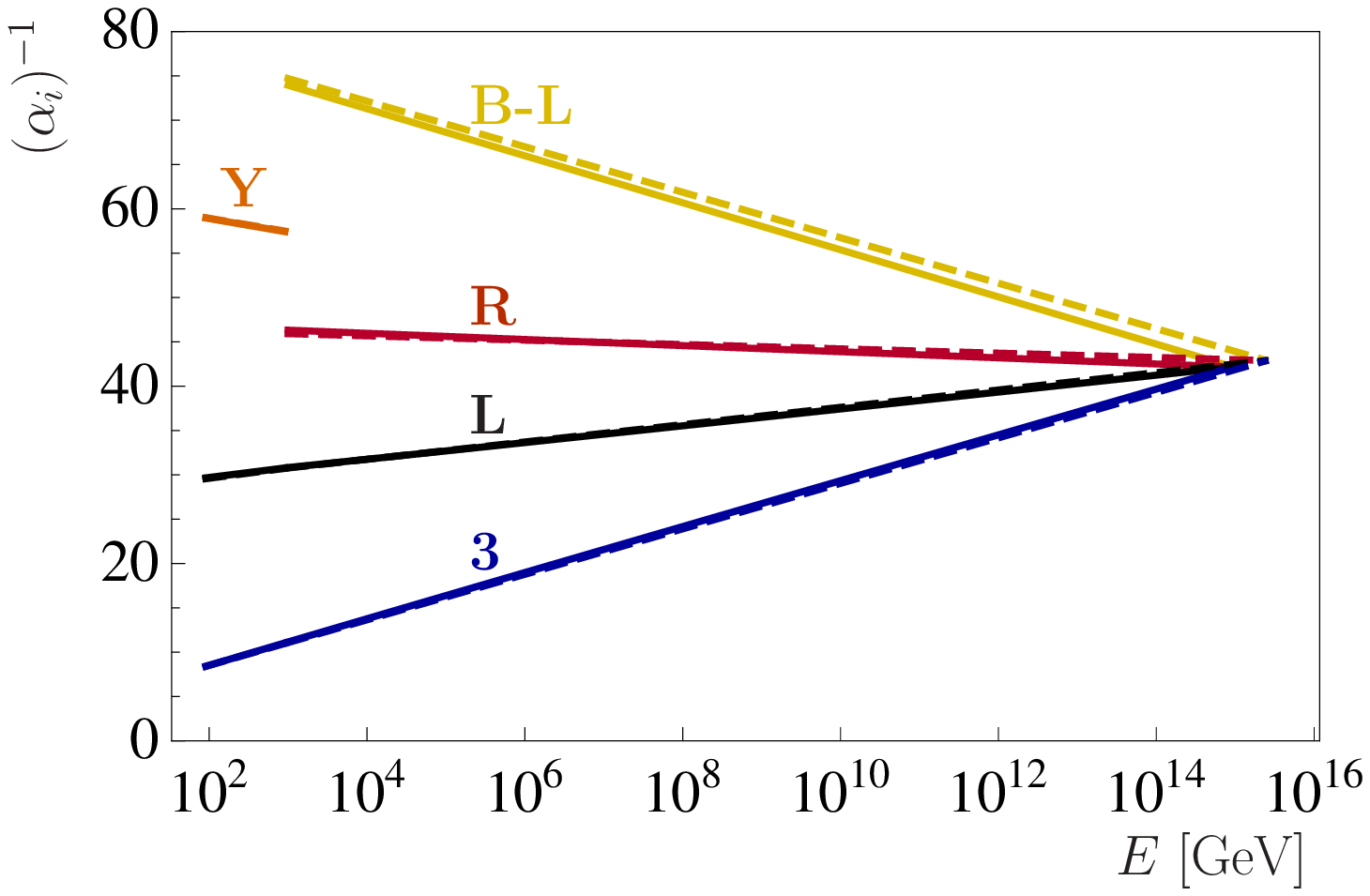}
\hskip-2mm\includegraphics[width=0.50\linewidth]{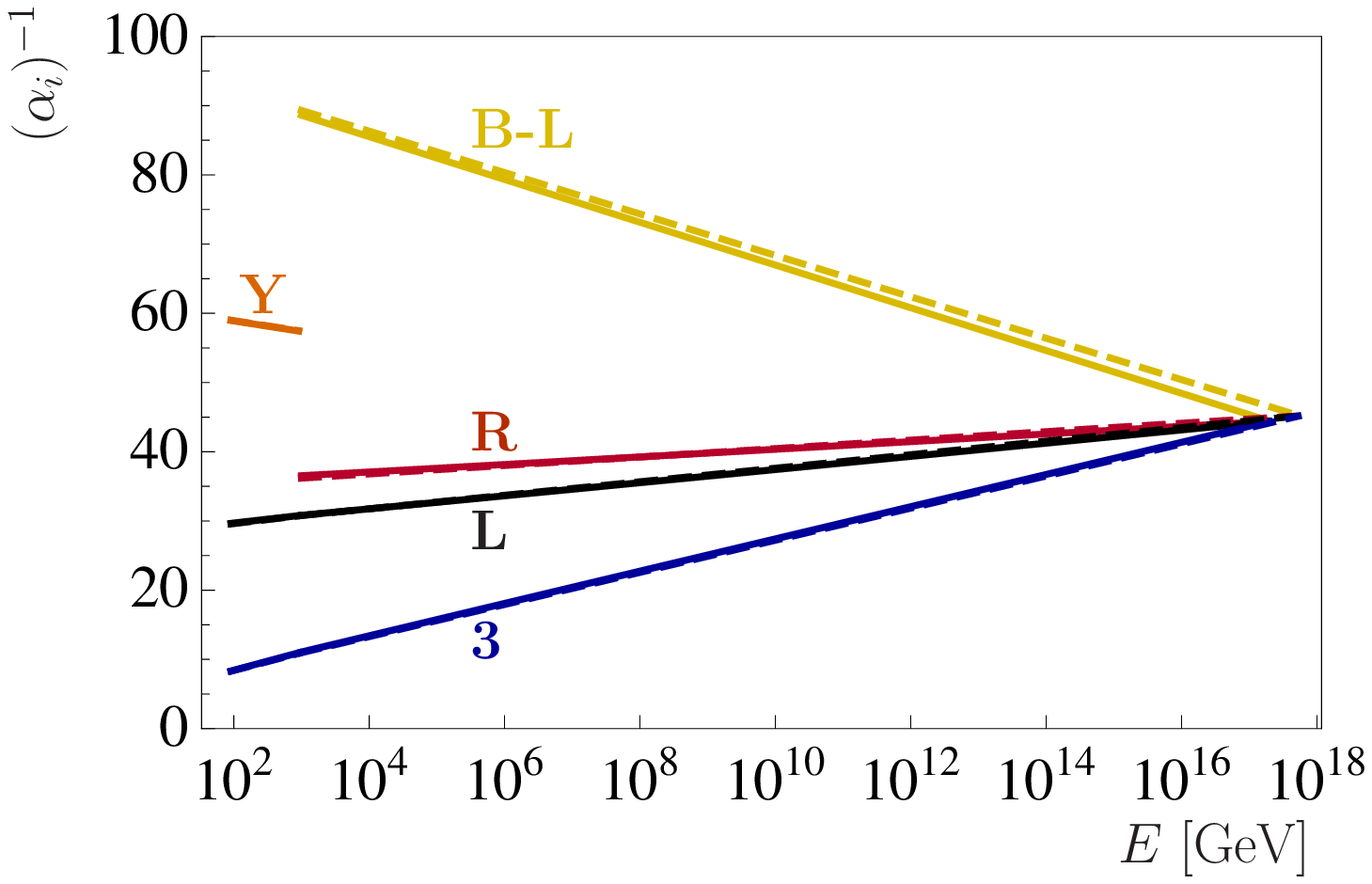}
\caption{Gauge coupling unification at 2-loop level 
(full lines) and 1-loop level (dashed lines), for two LR models 
with a low scale of LR breaking. The figure to the left has 
the field content SM + $\Phi_{1,2,2,0}+3 \Phi_{1,1,3,0}
+2\Phi_{1,1,3,-2}$, while the model to the right is defined 
as SM + $2\Psi_{3,1,1,-2/3}+2 \Phi_{1,2,1,1}
+2\Phi_{1,1,3,-2}$. For discussion see text.}
\label{fig:LRlow}
\end{figure}

Obviously, such a large scale for the LR-symmetry will never be probed
experimentally and this explains, perhaps, why LR models have not been
studied very much in the literature in the context of GCU. It is,
however, quite straightforward to construct LR symmetric models, where
the LR is close to the EW scale.  Just to give an indication, the
running of the inverse gauge couplings for two example models, which
we will discuss later in this paper and which lead to correct GCU with
a very low LR scale, are shown in fig.~\ref{fig:LRlow}. As discussed
in section~\ref{sec:LR}, many such examples can be constructed and
moreover, many of these examples give perfect GCU at a price of only
(a few copies of) one or a few additional types of fields.

Our work is, of course, not the first paper in the literature to
discuss GCU with a low LR scale. Especially supersymmetric models with
an extended gauge group have attracted recently some attention.
Different from the non-SUSY case, in SUSY LR models one needs to pay
special attention not to destroy the unification already achieved
within the MSSM. This can be done in different ways. In the
supersymmetric model of \cite{Malinsky:2005bi} the LR symmetry is
broken at a large scale, but the subgroup $U(1)_R\times U(1)_{B-L}$
survives down to the EW scale. In this construction, the scale where
$U(1)_R\times U(1)_{B-L}$ is broken to $U(1)_Y$ does not enter in the
determination of the GUT scale, $m_G$. Following
\cite{DeRomeri:2011ie} we will call such models "sliding scale"
models, since $U(1)_R\times U(1)_{B-L}$ can slide down from (nearly)
$m_G$ to any arbitrary value, without destroying GCU.  Also
supersymmetric sliding models with a full low-scale LR group can be
constructed, as shown in \cite{Majee:2007uv,Dev:2009aw}.
Alternatively, one can obtain sliding LR models, using an additional
intermediate scale, as has been shown in \cite{DeRomeri:2011ie}.  Many
examples of such ``sliding-scale'' supersymmetric LR constructions
have then be discussed in \cite{Arbelaez:2013hr}.

However, supersymmetry is not needed in low scale LR models to 
achieve GCU, as first discussed in the relatively unknown 
paper \cite{Lindner:1996tf}. Our work is based on similar ideas 
as this earlier paper  \cite{Lindner:1996tf}, but differs in 
the following aspects from it: 
(a) We do not insist on manifest LR symmetry. 
While parity maintaining LR models are, of course, a perfectly valid 
possibility, they only form a subclass of all LR models. 
(b) 
The study~\cite{Lindner:1996tf} concentrated exclusively on GCU. We also 
discuss constraints on model building due to the requirement 
of explaining correctly the CKM in LR symmetric models. We further
take in account constraints coming from the requirement that we should
have the necessary fields to have a successful seesaw mechanism for
neutrino masses.
(c) We add a discussion of ``sliding models''; as discussed above 
a particular (but interesting) sub-class of LR models.
And, 
(d) we pay special attention to uncertainties in the predictions 
of the LR and GUT scales (and the resulting uncertainty in the 
proton decay half-lives). As shown below, these uncertainties are 
entirely dominated by the current theory error, due to the (calculable 
but) unknown threshold errors.

The rest of this paper is organized as follows. In the next section we
discuss our minimal requirements for the construction of low-scale LR
symmetric models. Special emphasis is put on the discussion of how to
generate a realistic CKM matrix at tree-level. In section~\ref{sec:LR}
 we then discuss a number of possible LR models. We first consider
``minimal'' low-scale setups, i.e. models which fulfil all
requirements discussed in section~\ref{sec:requirements}
 with a field content as small as possible. We then discuss also
``sliding-scale LR models''. By this term we understand models, which
lead to the correct unification, but in which the scale, where LR
symmetry is broken, is essentially a free parameter.  This latter
models are non-minimal, but reminiscent of the supersymmetric LR
constructions discussed in \cite{Arbelaez:2013hr}.  In
section~\ref{sec:ChiSq} we then discuss uncertainties for the
prediction of the LR scale and the proton decay half-life in the
different models, before turning to a short summary and conclusion in
section~\ref{sec:conclusions}.  A number of details and tables of
possible models are given in the appendices.

\section{Basic requirements}
\label{sec:requirements}

There are several basic conceptual and phenomenological requirements
that we shall impose on the set of all possible LR-symmetric
extensions of the Standard model. From the bottom-up perspective these
are:
\begin{itemize}
\item {Rich enough structure to account for the CKM mixing even after
  the SM Higgs doublet is promoted to the LR bi-doublet, and a rich
  enough structure to support some variant of the seesaw mechanism.}
\item{Consistency of the assumed high-scale grand unified picture;
  here we shall be concerned, namely, with the perturbativity of the
  models up to at least the unification scale, the quality of the
  gauge coupling convergence (to be at least as good as in the minimal
  supersymmetric standard model) and compatibility with the current
  proton decay limits.}
\end{itemize}
Technically, we shall also assume that the masses of the extra degrees
of freedom are well clustered around at most two scales, i.e., the LR
scale and the GUT scale; if this was not the case there would be no
way to navigate through the plethora of possible
scenarios. Implicitly, the LR scale will be located in the TeV
ballpark otherwise decoupling would make the new physics escape all
LHC tests.

\subsection{Account for the SM flavour physics}
The need to accommodate flavour physics is clearly the least
speculative of the requirements above and, thus, the one we begin
with.

\subsubsection{Two bi-doublets plus one extra scalar}

With just the SM fermions at hand, there must obviously be more than a
single bi-doublet coupled to the quark and lepton bilinears in any
renormalizable LR-symmetric theory; otherwise, the Yukawa lagrangian
(in the ``classical'' LR notation with $Q\equiv \Psi_{3,2,1,{1}/{3}}$,
$\Phi\equiv \Phi_{1,2,2,0}$ and so on, cf. 
table (\ref{tab:List_of_LR_fields})) 
\be\label{single} {\cal
  L}_{Y}=Y_{Q}Q^{T}i\tau_{2}\Phi Q^{c}+Y_{L}L^{T}i\tau_{2}\Phi
L^{c}+h.c.\,, \ee yields $M_{u}\propto M_{d}$ irrespective of the
vacuum expectation value (VEV)
structure of $\Phi$ and, hence, $V_{CKM}=\mathbbm{1}$ at the
$SU(2)_{R}$ breaking scale.  With a second bi-doublet at play, one has
instead 
\be\label{Yukawa} 
{\cal L}_{Y}=Y_{Q}^{1}Q^{T}i\tau_{2}\Phi^{1}
Q^{c}+Y_{Q}^{2}Q^{T}i\tau_{2}\Phi^{2}
Q^{c}+Y_{L}^{1}L^{T}i\tau_{2}\Phi^{1}
L^{c}+Y_{L}^{2}L^{T}i\tau_{2}\Phi^{2} L^{c}+h.c.\,, 
\ee 
which admits $M_{u}$ non-proportional to $M_{d}$ (and, therefore, a
potentially realistic CKM provided\footnote{ Note that in the
  opposite case one can go into a basis in which one of the two
  bi-doublets is entirely deprived of its VEVs and, hence, one is
  effectively back to the single-$\Phi$ case (\ref{single}).}
\be\label{necessary} \frac{v_{u}^{1}}{v_{u}^{2}}\neq
\frac{v_{d}^{1}}{v_{d}^{2}}\,,\quad \text{where}\quad
\langle\Phi^{i}\rangle\equiv \begin{pmatrix} v_{d}^{i} & 0 \\ 0 &
  v_{u}^{i}
\end{pmatrix}\,.
\ee 
Note that we conveniently chose the $SU(2)_{R}$ index to label columns
(i.e., they change in the vertical direction) while the $SU(2)_{L}$
indices label the rows.

Needless to say, the VEV structure of such a theory is driven by the
relevant scalar potential. With just the two bi-doublets at play it
can be written in a very compact form \be\label{scalarsimplest}
V\ni-\frac{1}{2}\mu^{2}_{ij}\Tr(\tau_{2}\Phi^{iT}\tau_{2}\Phi^{j})\,,
\ee where the mass matrix $\mu$ can be, without loss of generality,
taken symmetric, cf. eq.~(6) in~\cite{Aulakh:1997ba}. In such a simple
case, however, it is almost obvious that the condition
(\ref{necessary}) can not be satisfied because of the
$\Phi^{1}\leftrightarrow \Phi^{2}$ interchange symmetry which yields
$v_{d}^{1}/v_{u}^{1}=v_{d}^{2}/v_{u}^{2}$ implying
$v_{d}^{1}/v_{d}^{2}=v_{u}^{1}/v_{u}^{2}$. Hence, either
eq.~(\ref{Yukawa}) or eq.~(\ref{scalarsimplest}) require further
ingredients.

Let us first try to devise (\ref{necessary}) by adding some extra
scalar fields so that the simple scalar potential
(\ref{scalarsimplest}) loses the $\Phi^{1}\leftrightarrow \Phi^{2}$
symmetry.
 
To this end, it is clear that the desired asymmetric term must contain
at least a pair of $\Phi$'s and anything that can be coupled to such a
bilinear, i.e., an $SU(2)_{R}$ singlet or a triplet, either elementary
(with a super-renormalizable coupling) or as a compound of two
doublets. Clearly, a singlet field (of any kind) behaves just like the
explicit singlet mass term in (\ref{scalarsimplest}) and, as such, it
does not lift the undesired degeneracy.

Hence, only the triplet option is viable, either in the form of an
elementary scalar\footnote{We discard the ``symmetric solution'' with
  an elementary ${\Omega}\equiv \Phi_{1,3,1,0}$ because such a field
  can not get any significant VEV without ruining the SM $\rho$
  parameter.} $\Phi_{1,1,3,0}$ (to be denoted $\Omega^{c}$, see again
table~\ref{tab:List_of_LR_fields}) which couples to the bi-doublets
via an {\em antisymmetric} coupling $\alpha$
\be\label{elementarytriplet} V\ni
\alpha_{ij}\Tr[\Phi^{iT}\tau_{2}\vec{\tau}\Phi^{j}\tau_{2}].\vec{\Omega}^{c}\,,
\ee or a non-elementary triplet made of a pair of $SU(2)_{R}$ doublets
$\chi^{c}$ $\equiv \Psi_{1,1,2,-1}$ (and $\chi^{c\dagger}$) replacing,
effectively, $\vec{\Omega}^{c}\to \chi^{c\dagger}\vec{\tau}\chi^{c}$.
Let us mention that the former option has been entertained heavily in
the SUSY LR context~\cite{Aulakh:1997ba,Aulakh:1997fq} where the
requirement of renormalizability of the superpotential simply enforces
this route; in the non-SUSY framework, however, the doublet solution
is at least as good as the triplet one.

To conclude, we shall consider all settings with the SM matter
content, a pair of LR bi-doublets and either and extra
$\Omega^{c}$-like $SU(2)_{R}$ triplet or an extra $\chi^{c}$-like
$SU(2)_{R}$ doublet consistent with the requirement of a realistic SM
flavour.

\subsubsection{Extra fermions}
Relaxing the strictly SM-like-matter assumption, one may attempt to
exploit the mixing of the chiral matter with possible vector-like
fermions emerging in various extensions of the SM. Among these, one
may, for instance, arrange the mixing of the SM left-handed quark
doublet $Q= \Psi_{3,2,+{1}/{6}}$ with the $Q'$ part of an extra
$Q$-type vector-like pair
\begin{eqnarray}
Q'\oplus Q'^{*}&\equiv & \Psi'_{3,2,+{1}/{6}}\oplus \Psi'_{\overline{3},\overline{2},-{1}/{6}}\,,
\end{eqnarray}
or a mixing of the SM $u^{c}=\Psi_{\overline{3},1,-{2}/{3}}$ and/or $d^{c}=\Psi_{\overline{3},1,+{1}/{3}}$ (in the notation in which all matter fields are left-handed) with the extra $u^{c}$ and/or $d^{c}$-like fields
\begin{equation}
{u'^{c}}\oplus {u'^{c*}}\equiv  \Psi'_{\overline{3},1,-{2}/{3}}\oplus \Psi'_{{3},1,+{2}/{3}}\,,\qquad 
{d'^{c}}\oplus {d'^{c*}}\equiv  \Psi'_{\overline{3},1,+{1}/{3}}\oplus \Psi'_{{3},1,-{1}/{3}}\,.
\end{equation}
For the sake of simplicity, we shall consider all these possibilities
at once and then focus on several special cases with either some of
these fields missing or with extra correlations implied by the
restoration of the LR symmetry at some scale.

The relevant piece of the Yukawa-type + mass lagrangian in such a case
reads (omitting all the gauge indices as well as the omnipresent
transposition and $C^{-1}$ Lorentz factors in all terms):
\begin{eqnarray}
{\cal L}_{Y+mass}^{matter}&=&Y_{u}Qu^{c}H_{u}+Y_{d}Qd^{c}H_{d}+Y'_{u}Q'u^{c}H_{u}+Y'_{d}Q'd^{c}H_{d}+Y'^{c}_{u}Qu'^{c}H_{u}+Y'^{c}_{d}Qd'^{c}H_{d}\nn\\
&+&Y''^{c}_{u}Q'u'^{c}H_{u}+Y''^{c}_{d}Q'd'^{c}H_{d}+ M_{Q'Q'^{*}}Q'Q'^{*}+M_{d'^{c}d'^{c*}}d'^{c}d'^{c*}+M_{u'^{c}u'^{c*}}u'^{c}u'^{c*}\nn\\
&+& M_{QQ'^{*}}QQ'^{*}+M_{d^{c}d'^{c*}}d^{c}d'^{c*}+M_{u^{c}u'^{c*}}u^{c}u'^{c*}+h.c.\label{efflagrangian}
\end{eqnarray}
where $Y_{u}$ and $Y_{d}$ are the standard $3\times 3$ Yukawa matrices
of the SM; the dimensionalities of the other matrix couplings (primed
$Y$'s) and/or direct mass terms ($M$'s) should be obvious once the
number of each type of the extra matter multiplets is specified.

In the QCD$\otimes$QED phase, this structure gives rise to the
following pair of the up- and down-type quark mass matrices (the last
columns and rows indicate whether the relevant field comes from an
$SU(2)_{L}$ doublet or a singlet and, hence, justify the qualitative
structure of the mass matrix; note also that we display only one of
the off-diagonal blocks of the full Dirac matrices written in the Weyl
basis and we do not pay much attention to ${\cal O}(1)$ numerical
factors such as Clebsches and/or normalisation):
\renewcommand*{\arraystretch}{1.2}
\begin{equation}
\begin{array}{c|ccc|c}
M_u & u^{c} & u'^{c*} & u'^{c} & SU(2)\\
\hline
u & Y_{u} v_{u}& M_{QQ'^{*}} & Y'^{c}_{u} v_{u} & 2\\
u' & Y'_{u} v_{u}& M_{Q'Q'^{*}} & Y''^{c}_{u} v_{u}& 2\\
u'^{c*} & M_{u^{c}u'^{c*}}^{T} &  Y''^{cT}_{u} v_{u}&M_{u'^{c}u'^{c*}}^{T} & 1\\
\hline
SU(2)& 1 & 2 & 1 & 
\end{array}\qquad
\begin{array}{c|ccc|c}
M_{d}& d^{c} & d'^{c*} & d'^{c} & SU(2)\\
\hline
d & Y_{d} v_{d}& M_{QQ'^{*}} & Y'^{c}_{d} v_{d} & 2\\
d' & Y'_{d} v_{d}& M_{Q'Q'^{*}} & Y''^{c}_{d} v_{d}& 2\\
d'^{c*} & M_{d^{c}d'^{c*}}^{T} &  Y''^{cT}_{d} v_{d}&M_{d'^{c}d'^{c*}}^{T} & 1\\
\hline
SU(2)& 1 & 2 & 1 &
\end{array}
\end{equation}
Given this, there are several basic generic observations one can make: 
\begin{itemize}
\item{The spectrum of both these matrices always contains three
  ``light'' eigenvalues, i.e., those that are proportional to the
  $SU(2)_{L}$ breaking VEV. This, of course, provides a trivial
  consistency check of their structure.}
\item{Removing the second row+column in both $M_{u,d}$ (that
  corresponds to integrating out $Q'\oplus Q'^{*}$) and/or the third
  row+column in $M_{u}$ (and, thus, integrating out ${u'^{c}}\oplus
  {u'^{c*}}$) and/or the third row+column in $M_{d}$ (and thus
  integrating out ${d'^{c}}\oplus {d'^{c*}}$) the game is reduced to
  all the different cases discussed in many previous studies in the SM
  context.}
\item{There are several entries in $M_{u}$ and $M_{d}$ that are
  intercorrelated already at the SM level; yet stronger correlations
  can be expected if the effective lagrangian (\ref{efflagrangian})
  descends from a LR-symmetric scenario. For example, grouping
  ${u'^{c}}\oplus {u'^{c*}}$ and ${d'^{c}}\oplus {d'^{c*}}$ into
  $SU(2)_{R}$ doublets ${Q'^{c}}\oplus {Q'^{c*}}$ the degeneracy among
  $M_{u}$ and $M_{d}$ would be exact up to (model-dependent)
  $SU(2)_{R}$-breaking terms; in such a case the (dis-)similarity of
  the up and down quark spectra and mixing matrices depends on the
  details of the specific $SU(2)_{R}$-breaking mechanism which,
  obviously, will be able to smear such degeneracies (and, thus, open
  room for a potentially realistic spectra and the CKM matrix) only
  if the relevant VEV is comparable to (or larger than) the singlet
  mass terms therein. Note that here we implicitly assume that there
  is no other mechanism such as the one described in the previous
  section operating to our desire.}
\end{itemize}
Hence, if one wants to make use of the extra vector-like fermions in
order to account for a realistic SM quark masses and mixing in the LR
setting, such extra matter fields should be included at (or below) the
LR scale, otherwise they will effectively decouple. This is the second
route to the realistic SM flavour that we shall entertain in what
follows.

To conclude, without going into more details, we shall consider all
scenarios including some of the combinations of the extra matter
fields discussed above with masses at the LR scale eligible for the
subsequent renormalization group (RG) analysis. In this respect, it is
also worth stressing 
that there are many specific realisations of the structures above at
the LR level that differ namely by the origin of the desired
vector-like fermions therein and, thus, by the specific structure of
the effective mass matrices above. An interested reader is deferred to
section~\ref{sec:LR} where several examples are discussed in more
detail.

\subsubsection{Seesaw \& neutrino masses}
We also require there are fields in the model that may support some
variant of the seesaw mechanism, either ordinary or inverse/linear,
and, thus, provide Majorana masses for neutrinos. Technically, the
requirements are identical to those given in the previous SUSY study
\cite{Arbelaez:2013hr} so we shall just recapitulate them here: i) in
models where the LR symmetry is broken by $\Phi_{1,1,3,-2}$ one
automatically has a right-handed neutrino mass and, thus, type-I seesaw; if
$\Phi_{1,3,1,-2}$ is also present, type-II contribution to the seesaw
formula is likely. ii) as for the models with the LR breaking driven
by $\Phi_{1,1,2,-1}$ one may implement either an inverse
\cite{Mohapatra:1986bd} and/or linear
\cite{Akhmedov:1995ip,Akhmedov:1995vm}, seesaw if $\Psi_{1,1,1,0}$ is
present, or a variant of type-III seesaw if $\Psi_{1,3,1,0}$ and/or
$\Psi_{1,1,3,0}$ is available.
\subsection{Consistency of the high-scale grand unification}
\subsubsection{Perturbativity}
Since the analysis in the next sections relies heavily on perturbative
techniques we should make sure these are under control in all cases of
our interest.  In particular, one should assume that for all couplings
perturbativity is not violated at $m_{G}$ and below $m_{G}$ the same
holds for all the effective parameters of the low-energy theory.  To
this end we shall, as usual, adopt a very simplified approach assuming
that none of the gauge couplings explodes throughout the whole
``desert'' and, at the same time, the unified coupling does not
diverge right above the unification scale.
On top of that, a perturbative description does not make much (of a
quantitative) sense either even if the couplings are formally
perturbative up to $m_{G}$ (and the spectrum is compact) when some of
them diverge very close above $m_G$: in fact, the results would be
extremely sensitive to the matching scale selection because their
rapid just-above-$m_G$ growth is equivalent to large thresholds for
not-so-well chosen matching scale.\\
\subsubsection{Grand unification}
Technically, $m_{G}$ is best defined as the mass scale of the heavy
vector bosons governing the perturbative baryon number violating (BNV)
processes. At first approximation, this may be determined as the
energy at which the running gauge couplings in the $\overline{\rm MS}$
scheme converge to a point; from consistency, this is then assumed to
be the scale where the heavy part of the scalar and vector spectrum is
integrated in.

Needless to say, if accuracy is at stakes, this picture is vastly
oversimplified.  The main issue of such an approach is the lack of a
detailed information about the high-energy theory spectrum which, in
reality, may be spread over several orders of magnitude\footnote{Note
  that this, in fact, is rather typical for ``simple'' models which
  tend to suffer from the emergence of pseudo-Goldstone bosons
  associated to spontaneously broken accidental global symmetries,
  especially when there are several vastly different scales at play,
  cf.~\cite{Bertolini:2009es}.}. The ``threshold effects'' thus
generated can then significantly alter the na\"\i ve picture by as
much as a typical two-loop $\beta$-function contribution.

This makes it particularly difficult to get a good grip on the GUT
scale from a mere renormalization group equations (RGE) running - with
the thresholds at play the 
running gauge couplings in the ``usual'' schemes such as
$\overline{\rm MS}$ do not intersect at a point and the only way
$m_{G}$ may be accurately determined is, indeed, a thorough inspection
of the heavy spectrum, see, e.g.,~\cite{Bertolini:2013vta}. In this
respect, perhaps the best that may be done in the bottom-up approach
(in which, by definition, the shape of the heavy spectrum is ignored)
is to define $m_{G}$ by means of a $\chi^{2}$ optimisation based on an
educated guess of the relevant theory error,
cf. section~\ref{sec:ChiSq}.

Another issue which often hinders the determination of $M_{G}$ is the
proximity of the unification and Planck scales which usually makes it
impossible to neglect entirely the Planck-suppressed effective
operators, especially those that, in the broken phase, make the gauge
kinetic terms depart from their canonical form. In the canonical
basis, these then yield yet another source of out-of-control shifts in
the GUT-scale matching conditions, i.e. smear the single-point gauge
unification picture yet further, see for instance \cite{Calmet:2008df}
and references therein. A simple back-of-the-envelope calculation
reveals that in most cases such effects are again comparable to those
of the two-loop contributions in the gauge beta functions.
Furthermore, the real cut-off $\Lambda$ associated to the quantum
gravity effects may be further reduced below the Planck scale if the
number of propagating degrees of freedom above is very large,
cf.~\cite{Dvali:2007hz}.

Since none of these issues may be addressed without a thorough
analysis of the coupled system of the two-loop renormalization group
equations augmented with a detailed information about the high-scale
spectrum (and, possibly, even quantum gravity), in what follows we
shall consider a unification pattern to be fine if the effective
$\overline{\rm MS}$ running gauge couplings do converge to a small
region characterised by a certain ``radius'' in the ``$t-\alpha^{-1}$
plot'' (with ${2\pi}t\equiv \log({\mu}/{M_{Z})}$ and $\mu$ denoting
the $\overline{\rm MS}$ regularization scale). Note that, in practice,
we shall perform a $\chi^{2}$-analysis of the gauge coupling RG
evolution pattern with three essentially free parameters at play,
namely, $m_{LR}$ (denoting the LR-scale where the part of the spectrum
that restores the $SU(2)_{R}$ gauge symmetry is integrated in),
$m_{G}$ (the scale of the assumed intersection of the relevant
effective gauge couplings of the intermediate-scale LR model) and
$\alpha_{G}$ (the unified ``fine structure'' coupling); with these
three degrees of freedom, however, an ideal fit of all three SM
effective gauge couplings, i.e., $\alpha_{s}$, $\alpha_L$ and
$\alpha_{Y}$, is (almost) always achievable. Hence, we shall push the
$\chi^{2}$-analysis further in attempt to assess the role of the
theoretical uncertainties in the possible future determination of
these three parameters that may be obtained in several different ways,
cf. Sect.~\ref{sec:ChiSq}.\\
 
\subsubsection{Proton lifetime}
There are in general many ingredients entering the proton lifetime
predictions in the grand unification context with very different
impact on their quality and accuracy.  Barring the transition from the
hadronic matrix elements to the hard quark-level correlators (assumed
to be reasonably well under control by the methods of the lattice QCD
and/or chiral Lagrangian techniques), these are namely the masses of
the mediators underpinning the effective BNV operators. At the $d=6$
level, these are namely the notorious GUT-scale $X$ and $Y$ (and/or
$X'$ and $Y'$) gauge bosons, and also the three types of potentially
dangerous scalars $\Phi_{3,1,-1/3}$, $\Phi_{3,1,-4/3}$ and
$\Phi_{3,3,-1/3}$ (descending from the fields nr. 9, 10, 14 and 19 in
table~\ref{tab:List_of_LR_fields}) with direct Yukawa couplings to
matter. In both cases, the flavour structure of the relevant BNV
currents is the central issue that can hardly be ignored in any
dedicated proton lifetime analysis. From this point of view, the
gauge-driven $p$-decay is usually regarded to as being under a better
control because it depends only on the (unified) gauge coupling and a
set of {\em unitary} matrices encoding transitions from the defining
to the mass bases in the quark and lepton sectors (whose matrix
elements, barring cancellations, are typically ${\cal O}(1)$) while
the scalar BNV vertices are governed by the Yukawa couplings and,
thus, are often (unduly) expected to be suppressed for the processes
involving the first generation quarks and leptons. In either case, a
detailed study of the flavour structure of the BNV currents is far
beyond the scope of the current study; the best one can do then is to
assume conservatively the gauge channels' dominance and suppose that
the elements of the underlying unitary matrices are of order~1.

However, in theories with accidentally light (TeV-scale) states one
should not finish at the $d=6$ level but rather consider also $d>6$
BNV transitions that may be induced by such ``unusual'' scalars. To
this end, let us just note that the emergence of $d=7$ baryon number
violating operators has been recently discussed in some detail
in~\cite{Babu:2012vb} (see also~\cite{Weinberg:1980bf,Weldon:1980gi})
and a specific set of scalars (in particular, $\Phi_{3,2,1/6}$,
$\Phi_{3,2,7/6}$ and $\Phi_{3,1,2/3}$) underpinning such transitions
in $SO(10)$ GUTs has been identified. Nevertheless, in the relevant
graphs these fields are often accompanied by the ``usual'' $d=6$
scalars above and, thus, for acceptable $d=6$ transitions the $d=7$
BNV operators tend to be also suppressed so we shall not elaborate on
them any further.

Since neither these issues may be handled without a very detailed
analysis of a specific scenario, for the sake of the simple
classification of potentially viable settings intended for the next
section we shall stick to the leading order (i.e., $d=6$) purely gauge
transitions and implement the current SK constraint of $\tau_{p\to
  \pi^{0}e^{+}}\gtrsim10^{34}$ years~\cite{Abe:2013lua}. This will be
imposed through the simple phenomenological formula
\begin{equation}\label{eq:ProtonLifetime}
\Gamma_{p}\approx\alpha_{G}^{2}{m_{p}^{5}}/{m_{G}^{4}}\,,
\end{equation}
which, technically, provides a further input to the $\chi^{2}$
analysis in section~\ref{sec:ChiSq}. We shall also ignore all the
effects related to pulling the effective $d=6$ operators from $m_{G}$
down to the electroweak scale, see, e.g.,
\cite{Buras:1977yy,Ellis:1979hy,Wilczek:1979hc}.

\section{Low scale left-right models}
\label{sec:LR}

We will first discuss the simplest variants of models, i.e. 
those with one new energy scale, which we will denote by $m_{LR}$.
Later on we will also discuss the possibility to have a ``sliding'' LR
scale ``on top'' of a SM-group stage with extended particle
content. These latter models are slightly more complicated in their
construction than the minimal ones, but interesting since they are
reminiscent of the supersymmetric sliding models discussed in
\cite{DeRomeri:2011ie,Arbelaez:2013hr}.

Although all our models are inspired by $SO(10)$ unification, we do
not concern ourself with the first step of symmetry breaking,
i.e. $SO(10) \to SU(3)_c \times SU(2)_L\times SU(2)_R \times
U(1)_{B-L}$.  The interested reader is referred to, for example,
\cite{Dev:2009aw} or \cite{Malinsky:2005bi}. In the LR stage, we
consider a total of 24 different representations, as listed in table
(\ref{tab:List_of_LR_fields}). These fields give all representations
found in $SO(10)$ multiplets up to ${\bf 126}$ and we consider
multiplets up to ${\bf 126}$ simply because the right triplet,
$\Phi_{1,1,3,-2}$, which presents one of the two simplest
possibilities to break the LR group correctly, is $\Phi_{1,1,3,-2} \in
{\bf 126}$ in the $SO(10)$ stage.  Larger multiplets could be easily
included, but lead of course to more elaborate models. The
transformation properties of all our allowed multiplets under the LR
group are given in table~\ref{tab:List_of_LR_fields} of the appendix.

In this section, we will keep the discussion mostly at the 1-loop
level for simplicity. Two-loop $\beta$-coefficients can be easily
included, but do not lead to any fundamental changes in the models
constructed.  Recall that at 1-loop order two copies of a complex
scalar give the same shift in the $\beta$-coefficients $\Delta(b_i)$
as one copy of a Weyl fermion. The coefficients for scalars and
fermions differ at two-loop order, of course, but these differences
are too small to be of any relevance in our model constructions
considering current uncertainties, see section~\ref{sec:ChiSq}.

\subsection{``Minimal'' models}
\label{subsect:MinLR}

The master equation for the running of the inverse gauge 
couplings at the 1-loop level can be written as:
\begin{equation}
\alpha^{-1}_i(t) = \alpha^{-1}_i(t_0) + \frac{b_i}{2 \pi}(t-t_0),
\label{eq:alpI}
\end{equation}
where $t_i = \log(m_i)$, as usual. The corresponding
$\beta$-coefficients are:
\begin{eqnarray}
\left(b_{3}^{SM},b_{2}^{SM},b_{1}^{SM}\right) & = & 
\left( -7, -19/6, 41/10 \right),
\nonumber \\
\left(b_{3}^{LR},b_{2}^{LR},b_{R}^{LR},b_{B-L}^{LR}\right) & = & 
\left (-7, -3, -3, 4\right)+\left(\Delta b_{3}^{LR},\Delta b_{2}^{LR},
\Delta b_{R}^{LR},\Delta b_{B-L}^{LR}\right).
\label{eq:bi}
\end{eqnarray}

The (B-L) charges in eq. (\ref{eq:bi}) are written in canonical
normalization.  Here, $\Delta b_{i}^{LR}$ stand for the contributions
from additional fields, not accounted for in the SM, while the
coefficients for the groups $SU(2)_L\times SU(2)_R$ include the
contribution from one bi-doublet field, $\Phi_{1,2,2,0}$. We decided
to include this field in the $b^{LR}_i$ directly, since the SM Higgs
$h = \Phi_{1,2,1/2} \in \Phi_{1,2,2,0}$ in all our constructions.

Next, $\alpha_{R}^{-1}(m_{LR})$ and $\alpha_{B-L}^{-1}(m_{LR})$ are
related to the SM hypercharge via: 
\begin{equation}
\alpha_{1}^{-1}(m_{LR})=
\frac{3}{5}\alpha_{R}^{-1}(m_{LR})+\frac{2}{5}\alpha_{B-L}^{-1}(m_{LR}).
\label{eq:match} 
\end{equation} 
Eq. (\ref{eq:match}) can be used to eliminate one of the four running
couplings from the system of equations, since the orthogonal
combination
$-\frac{2}{5}\alpha_{R}^{-1}(m_{LR})+\frac{3}{5}\alpha_{B-L}^{-1}(m_{LR})$
is a free paramter. Defining $\alpha^{eff}_{1}$, with a $\beta$ 
coefficient $\frac{3}{5}b_{R}^{LR}+\frac{2}{5}b_{B-L}^{LR}$ then allows 
finding the GUT scale using only three running couplings. 

\begin{figure}[t]
\includegraphics[width=0.50\linewidth]{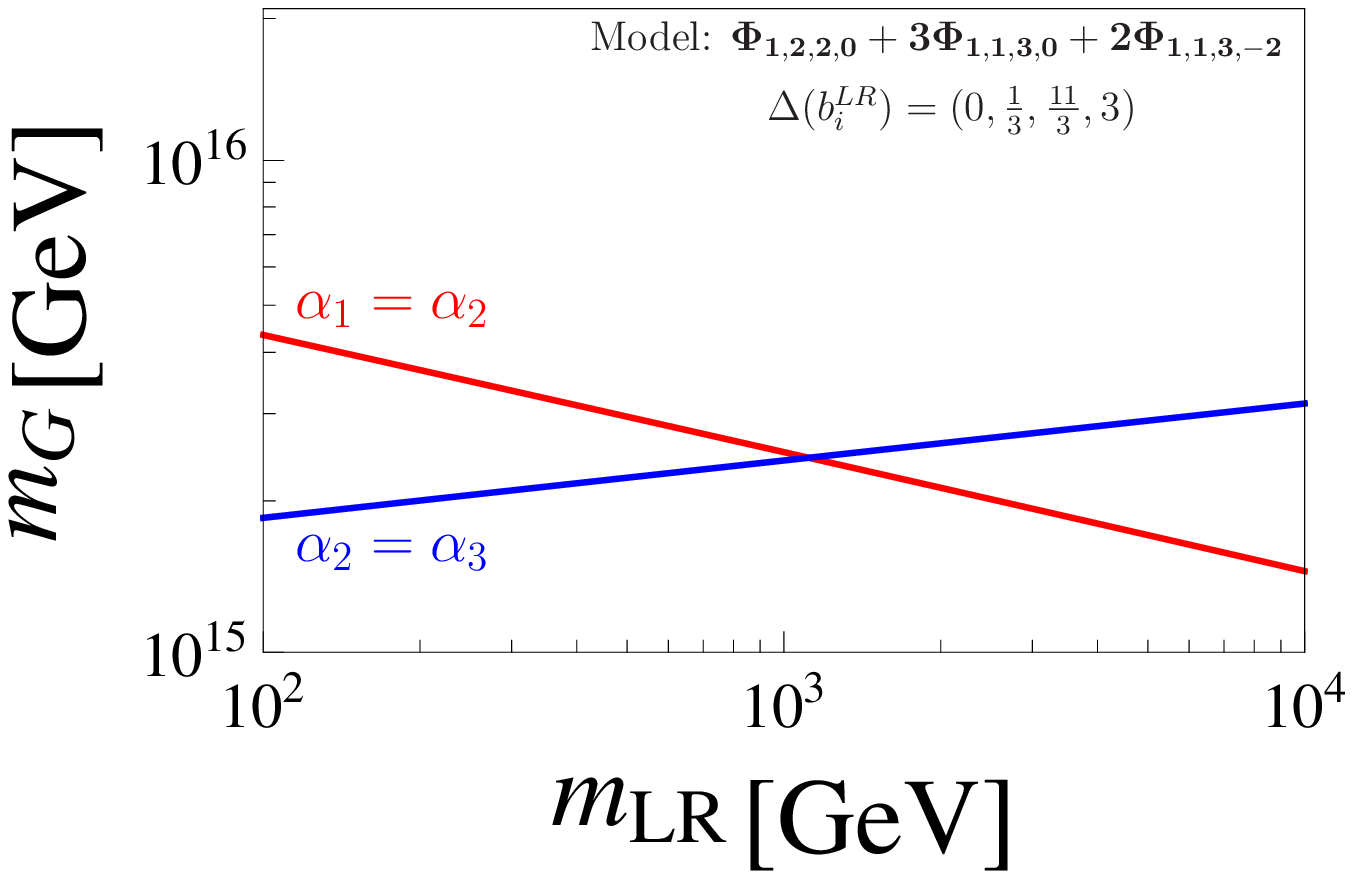}
\hskip-2mm\includegraphics[width=0.50\linewidth]{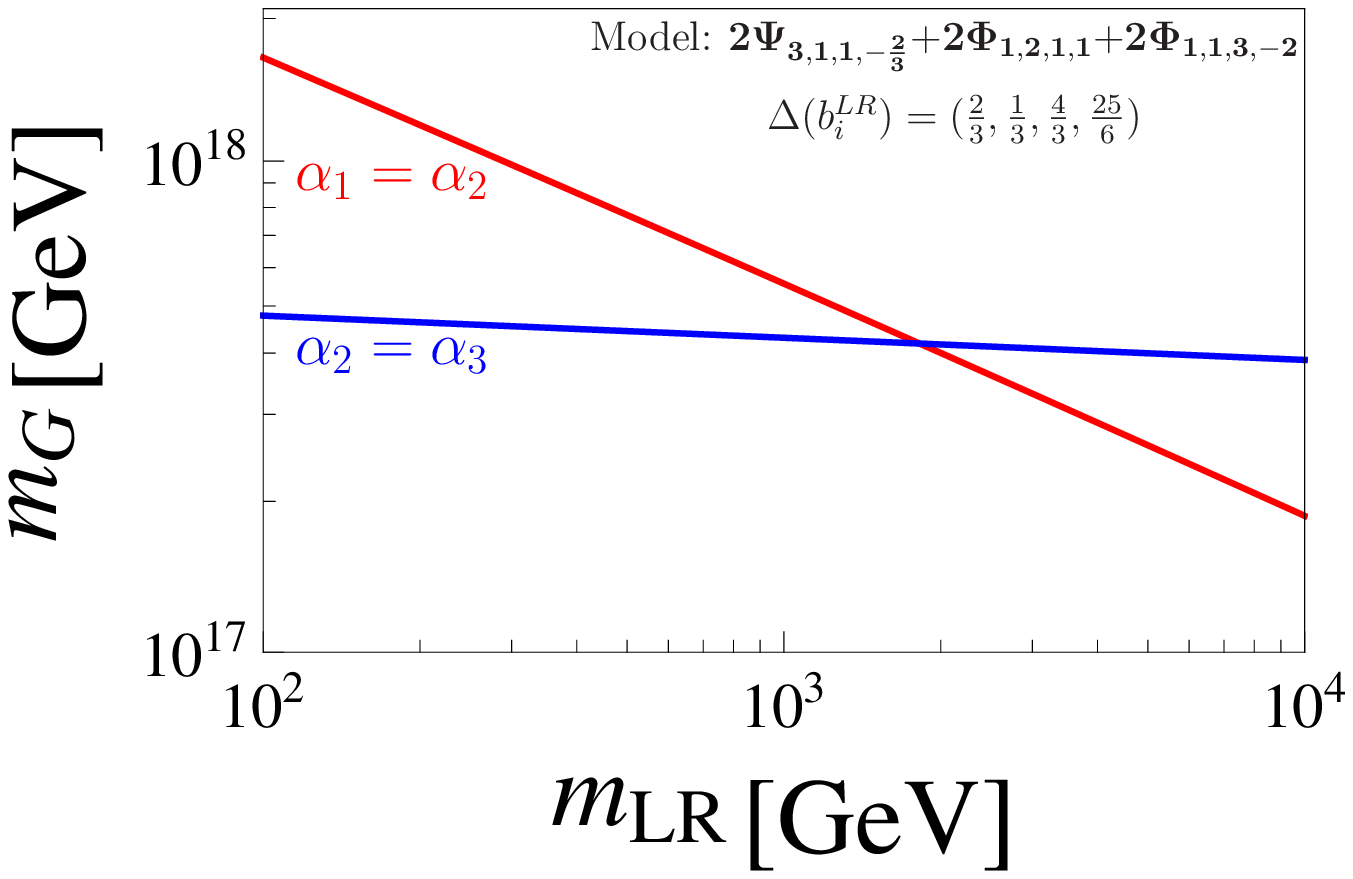}
\caption{Two example models for which correct unification is found for
  a low value of the scale $m_{LR}$. The model to the left has a
  rather low unification scale, see text. Note that these are the same
  two models already shown in fig.~\ref{fig:LRlow} in the
  introduction.}
\label{fig:Uni}
\end{figure}

Finding a model which unifies correctly, then simply amounts to
calculating a set of consistency conditions on the $\Delta(b^{LR}_i)$,
which can be derived from eq. (\ref{eq:alpI}), by equating
$\alpha_1^{eff}=\alpha_2$ and $\alpha_2=\alpha_3$. Two examples, for
which a correct unification is found with a low value of $m_{LR}$ are
shown in fig.~\ref{fig:Uni}. Note that, the model to the left has a
rather low unification scale (while the one to the right has a rather
high one). The half-life for proton decay in the best fit point at
1-loop level (at 2-loop level) for the model on the left is estimated
to be $T_{1/2} \simeq 10^{33}$ y ($T_{1/2} \simeq 10^{31}$ y), below
the lower limit from Super-K \cite{Nishino:2012ipa,Abe:2013lua}. This
will be important in the discussion on the error bar for proton decay
in section~\ref{sec:ChiSq} and is a particular feature of all model 
constructions without additional coloured fields, see below.

As discussed in the previous section, we then require a number of
additional conditions for a model to be both, realistic and
phenomenologically interesting: (i) All models must have the agents to
break the LR symmetry to the SM group; (ii) all models must contain
(at least) one of the minimal ingredients to generate a realistic CKM
and generate neutrino masses and angles; (iii) models must have
perturbative gauge couplings all the way to $m_G$; (iv) $m_G$ should be
large enough to prevent too rapid proton decay, numerically we have
used (somewhat arbitrarily) $m_G \ge 10^{15}$ GeV as the cut-off in
our search; and, lastly (v) the predicted $m_{LR}$ should be low
enough such that at least some of the new fields have masses accessible
at the LHC. As the cut-off in the search we used, again somewhat
arbitrarily, $m_{LR}=10$ TeV. \footnote{For both, $m_G$ and $m_{LR}$
  the values quoted are only the limits used in the search for
  models. Whether a particular model survives the constraints from
  proton decay searches depends not only on the values of $m_G$ and
  $\alpha_G$ but also on their uncertainties, see
  section~\ref{sec:ChiSq}.}   

Before discussing the different model classes, we first ask the
question how involved our constructions are. Different criteria can
be defined for comparing the complexity of different models, perhaps
the two simplest ones are: (i) $n_f$: the number of additional
different kinds of fields introduced and (ii) $n_c$: the total number
of new fields introduced. Consider first the classical, ``minimal''
high-scale LR models, mentioned already in the introduction. As shown
in table~\ref{tab:comp} the mLR
\cite{Mohapatra:1977mj,Mohapatra:1979ia,Mohapatra:1980yp} introduces
only 2 kind of fields, each with only one copy for a total of 2 new
fields, while the m$\Omega$LR already needs 5 different fields.
However, a realistic model should not only try to minimize the number
of new fields, it should also fulfil basic phenomenological
constraints discussed previously.  On this account, we would not
consider the mLR a valid model, since it has a trivial CKM at
tree-level, while the m$\Omega$LR is excluded (or at least at the
boundary of being excluded \footnote{See the discussion in 
section~\ref{sec:ChiSq}.}) by the constraints from the proton decay
half-life. The model mm$\Omega$LR (more-minimal $\Omega$LR), on the
other hand, can pass the phenomenological tests, with only
($n_f$,$n_c$)=(3,3). However, this model does not have $g_L=g_R$
(``exact parity'') at the scale where the LR symmetry is broken and
exact parity symmetry was required in most constructions of LR models,
that we have found in the literature. The question whether exact
parity (``manifest'') LR symmetry is a more important requirement for
a ``good'' model than having the smallest possible number of new
fields clearly is more a matter of taste than a scientific measure. We
decided not to insist on exact parity and instead construct models
with the fewest number of total fields possible.

The models we construct then can be separated into two different
classes: (a) models in which a realistic CKM is generated by the
extension of the scalar sector and (b) models in which a realistic CKM
is generated by the extension of the fermion sector.  

\subsubsection{Model class [a]: ``Scalar'' CKM models}

Consider first models of class (a). The breaking the LR group can be
either achieved via a right triplet, $\Phi_{1,1,3,-2}$ (case [a.1]),
or by a (right) doublet, $\Phi_{1,1,2,-1}$ (case [a.2]), as discussed
in the previous section. Several examples of simple models for both
classes are given in table~\ref{tab:comp}.

\begin{table}[t]
\begin{center}
\scalebox{0.95}{
\begin{tabular}{| l | l | l |  l | l |l | l |l |}
\hline
Name & Configuration & $n_f$ & $n_c$ & parity? & CKM? &  $m_{LR}$ [GeV] 
& $T_{1/2}$ [y] 
\\ \hline
mLR & $\Phi_{1,1,3,-2}+ \Phi_{1,3,1,-2}$ & 2 & 2 & \checked & \frownie & 
$3\cdot 10^{10}$  & $10^{33 \pm 2.5}$ \\ \hline 
m$\Omega$LR & $\Phi_{1,2,2,0}+\Phi_{1,1,3,0}+ \Phi_{1,3,1,0}
+\Phi_{1,1,3,-2}+ \Phi_{1,3,1,-2}$ & 5 & 5 & \checked &  \checked & 
$3 \cdot 10^{11}$ & $10^{30.8 \pm 2.5}$ \\ \hline 
mm$\Omega$LR  & $\Phi_{1,2,2,0}+\Phi_{1,1,3,0}+ 
+\Phi_{1,1,3,-2}$ & 3 & 3 & \frownie &  \checked & $3 \cdot 10^{9}$ 
& $10^{34.3 \pm 2.5}$ \\ \hline 
\end{tabular}}

\bigskip

\scalebox{0.95}{
\begin{tabular}{ | l | l |  l | l |l | l |l |}
\hline
Configuration & $n_f$ & $n_c$ & parity? & CKM? &  $m_{LR}$ [GeV] 
& $T_{1/2}$ [y] 
\\ \hline
 $\Phi_{1,2,2,0}+ \Phi_{1,1,3,0}+ 3 \Phi_{1,1,3,-2}$ & 3 & 5 
& \frownie  & \checked &  $1 \cdot 10^{2}$ 
& $10^{30.6 \pm 2.5}$ \\ \hline 
 $\Phi_{1,2,2,0}+3 \Phi_{1,1,3,0}+ 2 \Phi_{1,1,3,-2}$ & 3 & 6 
& \frownie  & \checked &  $2 \cdot 10^{3}$ 
& $10^{31.3 \pm 2.5}$ \\ \hline 
 $2 \Phi_{1,2,2,0}+ \Phi_{1,1,3,0}+ \Phi_{8,1,1,0} + 2 \Phi_{1,1,3,-2}$ & 4 & 6 
& \frownie  & \checked &  $5 \cdot 10^{2}$ 
& $10^{41.3 \pm 2.5}$ \\ \hline 
 $3 \Phi_{1,2,2,0}+ \Phi_{1,1,3,0}+3 \Phi_{6,1,1,4/3} + 2 \Phi_{1,3,1,-2} 
+ \Phi_{3,1,2,-2}$ & 5 & 10 
& \checked  & \checked &  $4 \cdot 10^{2}$ 
& $10^{36.3 \pm 2.5}$ \\ \hline 
\end{tabular}}

\bigskip

\scalebox{0.95}{
\begin{tabular}{ | l | l |  l | l |l | l |l |}
\hline
Configuration & $n_f$ & $n_c$ & parity? & CKM? &  $m_{LR}$ [GeV] 
& $T_{1/2}$ [y] 
\\ \hline
 $\Phi_{1,2,2,0}+ 16 \Phi_{1,1,2,-1}$ & 2 & 17  
& \frownie  & \checked &  $1 \cdot 10^{4}$ 
& $10^{31.6 \pm 2.5}$ \\ \hline 
 $\Phi_{1,2,2,0}+  \Phi_{1,1,2,-1}+  3 \Phi_{1,1,3,-2}$ & 3 & 5  
& \frownie  & \checked &  $2 \cdot 10^{3}$ 
& $10^{31.3 \pm 2.5}$ \\ \hline 
 $\Phi_{1,2,2,0}+  \Phi_{1,1,2,-1}+   \Phi_{1,1,3,-2}+ \Phi_{3,1,3,-2/3}$ & 4 & 4  
& \frownie  & \checked &  $2 \cdot 10^{3}$ 
& ??? \\ \hline 
$2 \Phi_{1,2,2,0}+\Phi_{1,1,2,-1} +\Phi_{6,1,1,-4/3}+2 \Phi_{1,1,3,-2} $ & 4 & 6 
& \frownie  & \checked &  $1 \cdot 10^{2}$ 
& $10^{39.6 \pm 2.5}$ \\ \hline 
$\Phi_{1,2,2,0}+2 \Phi_{1,1,2,-1} +2 \Phi_{1,2,1,1}+\Phi_{8,1,1,0}
+10 \Phi_{1,1,1,2} $ & 5 & 16 
& \checked  & \checked &  $3 \cdot 10^{3}$ 
& $10^{41 \pm 2.5}$ \\ \hline 
\end{tabular}}

\end{center}
\caption{A comparison of some of the simplest possible LR
  models. Configuration gives the actual (extra) fields used in the
  model on top of the SM fields. $n_f$ stands for $\#$(fields) and
  counts how many different fields are used in the construction, while
  $n_c$ is $\#$(copies) and counts the total number of different
  copies of fields. ``Parity?'' gives whether a given model predicts
  $g_L=g_R$ and ``CKM?'' whether it has a non-trivial CKM matrix at
  tree-level, see the discussion in the previous section. $m_{LR}$
  gives the approximate best fit point (including 2-loop coefficients)
  for the scale of LR breaking, while $T_{1/2}$ [y] gives the
  estimated half-life for proton decay. The error bar quoted for
  $T_{1/2}$ is an estimation derived from the discussion in 
section~\ref{sec:ChiSq}. The first table gives ``minimal'' LR models for
  comparison: These models all have $m_{LR}$ far above the EW
  scale. The second table gives models with low predicted $m_{LR}$ and
  CKM generated by scalar triplets (model class [a.1]), while the 3rd
  gives model examples with CKM generated by right-doublets (model
  class [a.2]). For discussion see main text. The model containing the
  field $\Phi_{3,1,3,-2/3}$ does not give a proton decay half-life,
  since the scalar field $\Phi_{3,1,3,-2/3}$ can induce proton decay
  via an unknown Yukawa coupling.}
\label{tab:comp}
\end{table}

Consider the triplet case first. The minimal field content for the
triplet case consists in $n_{\Phi_{1,2,2,0}} \Phi_{1,2,2,0}+
n_{\Phi_{1,1,3,0}} \Phi_{1,1,3,0}+n_{\Phi_{1,1,3,-2}} \Phi_{1,1,3,-2}$
and the simplest model we have found is given by
$n_{\Phi_{1,2,2,0}}=1$, $n_{\Phi_{1,1,3,0}}=1$ and
$n_{\Phi_{1,1,3,-2}}=3$ for a total of $n_c=5$ copies, followed by
$n_{\Phi_{1,2,2,0}}=1$, $n_{\Phi_{1,1,3,0}}=3$ and
$n_{\Phi_{1,1,3,-2}}=2$ for a total of $n_c$=6. Both models have
rather short proton decay half-lives, with the $n_c$=6 model doing
slightly better than the $n_c$=5 model. For this reason we used the
$n_c$=6 model in figs (\ref{fig:LRlow}) and (\ref{fig:Uni}) and in
section~\ref{sec:ChiSq} for our discussion. Once additional new fields
are allowed with non-zero coefficients, a plethora of models in this
class can be found. Example models for each of the 24 fields 
are given in table~\ref{tab:a.1} in the appendix. Here,
let us only briefly mention two more examples: $2 \Phi_{1,2,2,0}+
\Phi_{1,1,3,0}+ \Phi_{8,1,1,0} + 2 \Phi_{1,1,3,-2}$ and $3
\Phi_{1,2,2,0}+ \Phi_{1,1,3,0}+3 \Phi_{6,1,1,4/3} + 2 \Phi_{1,3,1,-2}
+ \Phi_{1,1,3,-2}$. The former shows (see discussion of
fig.~\ref{fig:nocol} below) that at the price of introducing one
coloured field, the proton decay half-life constraint can be
completely evaded, while the latter demonstrates that it is possible
to obtain exact parity symmetry even with different number of copies
of fields in the left and right sector of the model - at a price of a
few additional copies of fields.

Consider now model class [a.2]: $n_{\Phi_{1,2,2,0}} \Phi_{1,2,2,0}+
n_{\Phi_{1,1,2,-1}} \Phi_{1,1,2,-1}+ \cdots$. In this case, in
principle the simplest model possible consists in only two different
fields, since $\Phi_{1,1,2,-1}$ can play the double role of breaking
the LR symmetry and generating the non-trivial CKM, as explained in
the previous section. However, as table~\ref{tab:comp} shows, our
condition of having a low $m_{LR} \lsim 10$ TeV enforces a large
number of copies for this possibility: $n_{\Phi_{1,2,2,0}}=1$, but
$n_{\Phi_{1,1,2,-1}}=16$, not a very minimal possibility. 
Table~\ref{tab:comp} also shows that with three different fields, much
smaller multiplicities lead to consistent solutions. With 3 different
fields a solution with $n_c$=5 exists, for four different fields
$n_c$=4 is possible in one example. However, again, the example with
$n_c$=5 has a rather short $T_{1/2}$, while the $n_c$=4 contains a
copy of $\Phi_{3,1,3,-2/3}$. This field induces proton decay via a
dimension-6 operator, see discussion in the previous section and thus
does not lead to a realistic model, unless either the $\Delta(L)=1$ or
the $\Delta(B)=1$ Yukawa coupling is eliminated by the imposition of 
some symmetry. The next simplest model then contains
($n_f$=4,$n_c$=5). This case, however, has a b.f.p. for the $m_{LR}$ 
above our usual cutoff. Once we allow for ($n_f$=4,$n_c$=6) or larger, 
again many possibilities exist, one example is given in table~\ref{tab:comp}.
As for the case [a.1], models with exact parity are possible, but require 
a larger number of copies of fields.

\begin{figure}[t]
\includegraphics[width=0.50\linewidth]{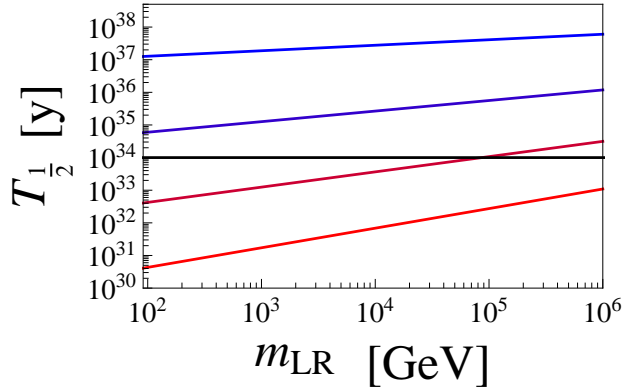}
\caption{One-loop estimated proton lifetime for
  ``colourless models'' as a function of $m_{LR}$. The figure shows
  $T_{1/2}$ [y] estimated from $m_G$ defined as the point where
  $\alpha_2=\alpha_3$ with from top to bottom: $\Delta(b^{LR}_2) = 0$,
  $\frac{1}{6}$, $\frac{1}{3}$ and $\frac{1}{2}$ and $\Delta(b^{LR}_3)
  = 0$ (``colourless models''), see text. The horizontal line is the
  experimental limit from Super-K \cite{Nishino:2012ipa,Abe:2013lua}.}
\label{fig:nocol}
\end{figure}  

Before closing this discussion on model class (a), we briefly comment
on the comparatively low values for the proton lifetime for
all cases in which {\em no coloured field} is added to the
configuration.  In the SM (with one Higgs and at 1-loop order)
$\alpha_2$ equals $\alpha_3$ at a scale of roughly $m_{G_{23}} = 10^{17}$
GeV. Adding a second Higgs, as necessary to complete the bi-doublet in
our LR models,\footnote{As in the MSSM, where a second Higgs doublet must
  be present.} lowers this GUT scale to roughly $m_{G_{23}}=2 \cdot
10^{16}$ GeV. Any addition of a field charged under $SU(2)_L$
increases $b_2$, leading to a further reduction in $m_{G_{23}}$,
unless some coloured field is added at the same time. Thus, all models
with a second $\Phi_{1,2,2,0}$ (or other fields charged under $SU(2)_L$)
but no additional coloured particles will have a GUT scale below
$10^{16}$ GeV. This is indeed quite an important constraint, as is
shown in fig.~\ref{fig:nocol}. Recall, for a $\Phi_{1,2,1,1}$ the
$\Delta(b^{LR}_2) = \frac{1}{6}$, while for $\Phi_{1,2,2,0}$ the
$\Delta(b^{LR}_2) = \frac{1}{3}$. Thus, ``colourless'' models can have
at most one additional $\Phi_{1,2,2,0}$, otherwise they are ruled out
by proton decay constraints. We note, that the figure is based on a 
1-loop calculation and that this conclusion is only
strengthened, once 2-loop $\beta$ coefficients are included, 
compare to the lifetimes quoted in table~\ref{tab:comp}.

\subsubsection{Model class [b]: ``Fermionic'' CKM models}

\begin{table}[t]
\begin{center}
\scalebox{0.95}{
\begin{tabular}{| l | l | l |  l | l |l | l |}
\hline
Configuration & $n_f$ & $n_c$ & parity? & CKM? &  $m_{LR}$ [GeV] 
& $T_{1/2}$ [y] 
\\ \hline
$2 \Psi_{3,1,1,-2/3}+ 2 \Phi_{1,2,1,1}+2\Phi_{1,1,3,-2}$ & 4 & 6 &
 \frownie & \checked 
&  $3\cdot 10^{3}$  & $10^{40 \pm 5}$ \\ \hline 
$2 \Psi_{3,1,1,-2/3}+ 2 \Phi_{1,1,2,-1}+  \Phi_{1,2,2,0}+  4 \Phi_{1,1,3,0}$ 
& 5 & 11 
& \frownie & \checked  &  $1\cdot 10^{4}$  & $10^{39.9 \pm 2.5}$ \\ \hline 
$2 \Psi_{3,1,1,-2/3}+ 2 \Phi_{1,1,2,-1}+ 2 \Phi_{1,2,1,1}+  4 \Phi_{1,1,3,0}$ 
& 5 & 12 
& \frownie & \checked  & $9\cdot 10^{3}$  & $10^{39.9 \pm 2.5}$ \\ \hline 
$2 \Psi_{3,1,1,-2/3}+\Phi_{1,2,1,1}+\Phi_{1,1,2,-1}+9\Phi_{1,1,1,2}$ & 5 & 13 
& \checked & \checked &  $1\cdot 10^{2}$  & $10^{43.4 \pm 2.5}$ \\ \hline 
$2 \Psi_{3,1,1,4/3}+ 3 \Phi_{1,2,1,1}+ \Phi_{1,1,3,-2}+  \Phi_{3,1,1,4/3}$ & 5 & 7 
& \frownie & \checked  & $6\cdot 10^{3}$  & $10^{40 \pm 2.5}$ \\ \hline 
$2 \Psi_{3,1,1,4/3}+ 3 \Phi_{1,2,1,1}+ 5 \Phi_{1,1,2,-1}+  \Phi_{3,1,1,4/3}$ 
& 5 & 11 
& \frownie & \checked  & $1\cdot 10^{4}$  & $10^{40 \pm 2.5}$ \\ \hline 
$2 \Psi_{3,2,1,1/3}+  \Phi_{8,1,1,0}+ 4 \Phi_{1,1,3,-2}$ & 4 & 7 
& \frownie & \checked  & $1\cdot 10^{2}$  & $10^{43 \pm 2.5}$ \\ \hline 
$2 \Psi_{3,2,1,1/3}+  \Phi_{6,1,1,2/3}+ 4 \Phi_{1,1,3,-2}$ & 4 & 7 
& \frownie & \checked  & $1\cdot 10^{2}$  & $10^{39.3 \pm 2.5}$ \\ \hline 
$2 \Psi_{3,2,1,1/3}+  \Psi_{3,1,3,-2/3}+ 6 \Phi_{1,1,3,-2}$ & 4 & 9
& \frownie & \checked  & $4\cdot 10^{3}$  & $10^{40.3 \pm 2.5}$ \\ \hline 
\end{tabular}}
\end{center}
\caption{A comparison of models with CKM generated by an extension in
  the fermion sector, ``fermionic CKM'' or ``VLQ-CKM''.  In $n_f$ we
  always count the two $\Psi_{3,i,j,k}$ as two separate fields,
  because both $\Psi$ and ${\bar\Psi}$ are needed to generate the
  CKM.}
\label{tab:vlqs}
\end{table}

We now turn to a discussion of models with additional fermions, see
table~\ref{tab:vlqs}. As discussed in
section~\ref{sec:requirements}, a non-trivial CKM can be generated in
LR models with extensions in the fermion sector essentially by three
kind of fields, corresponding to vector like copies of the SM fields
$u^c$, $d^c$ and $Q$. In the list of 24 different fields shown in
table~\ref{tab:List_of_LR_fields} in the appendix, there are in fact
several which contain states which can play the role of the VLQs after
the breaking of the LR symmetry.

Consider, for example, the case of $u^{'c} = \Psi'_{{\bar
    3},1,-2/3}$. The $\Psi'_{{\bar 3},1,-2/3}$ could be generated from
$\Psi_{{\bar 3},1,-2/3} \in {\bar \Psi}_{{\bar 3},1,1,-4/3}$,
${\bar\Psi}_{{\bar 3},1,2,-1/3}$ or ${\bar \Psi}_{{\bar 3},1,3,2/3}$.
Similarly, $d^{'c} = \Psi'_{{\bar 3},1,1/3} \in {\bar \Psi}_{{\bar
    3},1,1,2/3}$, ${\bar\Psi}_{{\bar 3},1,2,-1/3}$ or ${\bar
  \Psi}_{{\bar 3},1,3,2/3}$, while $Q' = \Psi'_{3,2,1/6} \in
\Psi_{3,2,1,1/3}$, $\Psi_{3,2,2,4/3}$, $\Psi_{3,3,1,-2/3}$ and
$\Psi_{3,2,2,-2/3}$. In the SM regime, therefore, different terms from
the LR regime can lead to the same effects.  We will consider only the
three simplest possibilities here, $\Psi'_{3,1,1,4/3}$,
$\Psi'_{3,1,1,-2/3}$ and $\Psi'_{3,2,1,1/3}$, where we have marked the
fields with a prime again to note that they have to be introduced in
vector-like pairs.  Other cases can be constructed in a similar
manner. For these three fields the corresponding Lagrangian terms in
the LR-regime are:
\begin{eqnarray}\label{eq:lagvlq}
{\cal L} & =& m_{\Psi_{3,1,1,4/3}}\Psi'_{3,1,1,4/3}{\bar \Psi}'_{{\bar 3},1,1,-4/3} 
         + m_{\Psi_{3,1,1,-2/3}}\Psi'_{3,1,1,-2/3}{\bar \Psi}'_{{\bar 3},1,1,2/3} 
\\ \nonumber
         &+ &m_{\Psi_{3,2,1,1/3}}\Psi'_{3,2,1,1/3}{\bar \Psi}'_{{\bar 3},2,1,-1/3} 
 + {\bar Y}_{\Psi_{3,1,1,4/3}}{\bar \Psi}'_{{\bar 3},1,1,-4/3}\Phi_{1,2,1,1}
   \Psi_{3,2,1,1/3}
\\ \nonumber
  & +& Y_{\Psi_{3,1,1,4/3}}\Psi_{3,1,1,4/3}'\Phi_{1,1,2,-1}\Psi_{{\bar 3},1,2,-1/3}
  + {\bar Y}_{\Psi_{3,1,1,-2/3}}{\bar \Psi}'_{{\bar 3},1,1,2/3}
    {\bar \Phi}_{1,2,1,-1}\Psi_{3,2,1,1/3}
\\ \nonumber
&+& Y_{\Psi_{3,1,1,-2/3}} \Psi'_{ 3,1,1,-2/3}
    {\bar \Phi}_{1,1,2,1}\Psi_{{\bar 3},1,2,-1/3}
   + Y_{\Psi_{3,2,1,1/3}}\Psi'_{3,2,1,1/3}\Phi_{1,2,2,0}\Psi_{{\bar 3},1,2,-1/3},
\end{eqnarray}
where $\Psi_{3,2,1,1/3} = Q$ and $\Psi_{{\bar 3},1,2,-1/3} =Q^c$ 
correspond to the SM left and right-handed quarks in the LR regime. 
Note, that $\Phi_{1,2,2,0}$ contains the SM-like VEV $v_u$, while 
for $\Psi_{3,1,1,4/3}$ and $\Psi_{3,1,1,-2/3}$ the corresponding mass 
terms are generated from the VEVs of $\Phi_{1,2,1,1}$ and $\Phi_{1,1,2,-1}$.
Recall that, as discussed in section~\ref{sec:requirements}, not all 
terms are necessary and in principle two terms (one mass term and one 
Yukawa term) are sufficient in all cases to generate the desired 
structure.

In table~\ref{tab:vlqs} we give some simple example models for these
cases: $\Psi'_{3,1,1,-2/3}$, $\Psi'_{3,1,1,4/3}$ and
$\Psi'_{3,2,1,1/3}$. Here, we wrote $2\Psi$ for $\Psi+{\bar\Psi}$
simply to get a more compact table. Since we count these as two
different kinds of fields and at least one $\Phi_{1,1,3,-2}$ or
$\Phi_{1,1,2,-1}$ is needed to break the LR symmetry, the minimal
$n_f$ seems to be three in these constructions. However, once we
impose $m_{LR}\lsim 10$ TeV, no solution with $n_f$=3 survives,
although there are many solutions with $n_f$=4 and 5. Perhaps the
simplest case possible is the model in the first line, which fulfils
all our conditions for the price of just two extra $\Phi_{1,1,2,1}$
and one extra $\Phi_{1,1,3,-2}$. In general, models which break the LR
symmetry via $\Phi_{1,1,2,-1}$ need more copies of fields to get a
consistent model with low $m_{LR}$, $n_c$$\ge 11$. Also, it is
possible to conserve parity exactly, as the table shows. However, the
model with the smallest $n_c$ that we found still has $n_c$=13. We
have not found any model with less than $n_c$=7 for the cases
$\Psi'_{3,1,1,4/3}\to u^{c'}$ and $\Psi'_{3,2,1,1/3}\to Q'$.

In case of models with VLQs, the constraints from proton decay are 
relatively easy to fulfil, see table~\ref{tab:vlqs}. This is simply 
due to the fact that VLQs add a non-zero $\Delta(b^{LR}_3)$, by which 
$m_G$ can be raised to essentially any number desired. 

\subsection{``Sliding'' LR models}
\label{subsect:SlidLR}

We now turn to the discussion of ``sliding-LR'' models. These are
defined as models where the unification is independent of the
intermediate scale $m_{LR}$. In (minimal) supersymmetric extensions of
the SM ``sliding-LR'' models are the only possibility to have a low
$m_{LR}$ \cite{Malinsky:2005bi,DeRomeri:2011ie,Arbelaez:2013hr}.
However, as we show in this subsection, supersymmetry is {\em not} a
necessary ingredient to construct sliding models.

We will discuss in the following just two examples of sliding LR
models. The first one, based on the idea of ``split'' supersymmetry
\cite{ArkaniHamed:2004fb,Giudice:2004tc}, shows the relation of our
non-supersymmetric sliding models, with the supersymmetric ones
discussed in \cite{Arbelaez:2013hr}. The second one is based on a SM
extension with vector-like quarks, first mentioned in
\cite{Amaldi:1991zx} and recently discussed in much more detail in
\cite{Gogoladze:2010in}. This second example serves to show, how non-SUSY
sliding models can be just as easily constructed as supersymmetric
ones.

\noindent 
The sliding conditions can be understood as a set of conditions on the
allowed $\beta$ coefficients of the gauge couplings in the LR regime
\cite{Arbelaez:2013hr}, assuring that at 1-loop order
$\Delta(\alpha_i)$ at the GUT scale are independent of the additional
particle content in the LR regime. In order to achieve successful
unification, therefore, it is necessary to first add to the standard
model an additional field content at some scale $m_{NP}$. Although not
necessary from a theoretical point of view, we require that $m_{NP}$ is
at a ``low'' scale, i.e. $m_{NP}\lsim 10$ TeV, to ensure that the
models predict some interesting collider phenomenology. We will call
this additional field content ``configuration-X'' and ``SM+X''.  A
list of simple X-configurations, which when added to the SM at
$m_{NP}$ in the range $m_{NP}$ (few) TeV lead to unification as
precise or better than the one obtained in the MSSM, is given in 
table~\ref{tab:X} in the appendix. In this table (at least) one example for
each one of our $24$ fields is presented.

\noindent  
As the first example, we will discuss the ``split SUSY-like'' case,
which corresponds to $X=5\Phi_{1,2,1/2}+2\Phi_{1,3,0}+2\Phi_{8,1,0}$.
As is well-known, in split SUSY the sparticle spectrum is "split" in
two regimes: all scalars (squarks, sleptons and all Higgs fields
except $h^0$) have masses at a rather high scale, typically $10^{10}$
GeV, while the fermions, gluino ($\Psi_{8,1,0}$), wino
($\Psi_{1,3,0}$), bino ($\Psi_{1,1,0}$) and the higgsinos
($\widetilde{H}_{u}=\Psi_{1,2,-1/2}$, and
$\widetilde{H}_{d}=\Psi_{1,2,1/2}$) must have TeV-ish masses. This way
GCU is maintained with a $\Delta(\alpha_i)$ at the GUT scale as small
as is the case in the MSSM (but at a different value of $\alpha_G$).
However, while in split SUSY $\Phi_{1,2,1/2}$ is added at the high
scale, for our LR constructions we will need this second Higgs at a low
scale and, therefore, we call this scenario ``split SUSY-like''. Note
that, while split SUSY uses fermions at the low scale, GCU can be
maintained also with a purely bosonic $X$, since only the 2-loop
coefficients change (slightly), which can be compensated by a slight
shift in $m_{NP}$.  We note in passing that this particular $X$ has,
of course, all the interesting phenomenology of split SUSY, like a
candidate for the dark matter, or a quasi-stable gluino at the LHC
\cite{ArkaniHamed:2004fb}.

\begin{figure}[htb]
\centering
\begin{tabular}{cc}
\includegraphics[width=0.49\linewidth]{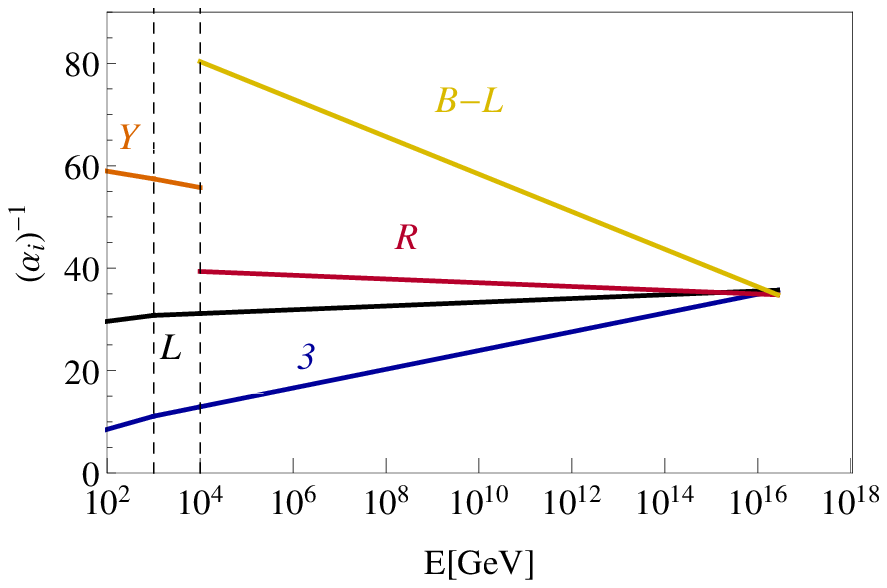}&
\includegraphics[width=0.49\linewidth]{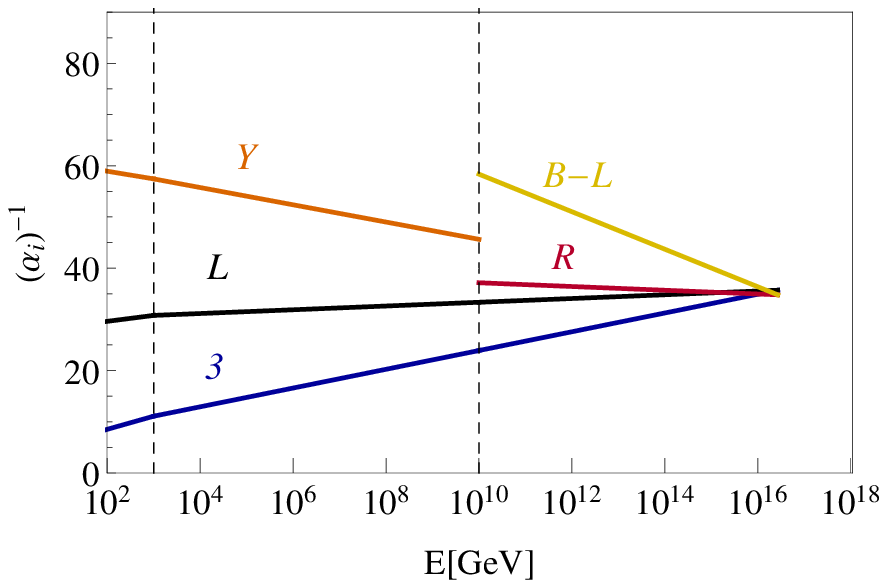}
\end{tabular}
\caption{Evolution of the gauge couplings for the sliding-LR model
  example discussed in the text based on split SUSY. The plot to the
  left shows $m_{LR}=10$ TeV, while the plot to the right has
  $m_{LR}=10^{10}$ GeV.}
\label{fig:slidingGCU}
\end{figure}

The quantum numbers of this particular particle content in the 
LR regime are then: $\Phi_{1,2,1/2}\in \Phi_{1,2,2,0}$, 
$\Phi_{1,3,0}\in \Phi_{1,3,1,0}$ and $\Phi_{8,1,0}\in \Phi_{8,1,1,0}$, 
with the $\Delta b^{LR}_{i}$ coefficients corresponding to this particular 
$X$ given by:
\begin{equation}
(\Delta b_{3}^{LR},\Delta b_{2}^{LR}, \Delta b_{R}^{LR}, \Delta b_{B-L}^{LR}) 
= (2,2,{2}/{3},0).
\end{equation}
Imposing now the requirement that $m_{G}$ is independent of the
intermediate scale $m_{LR}$, results in  the set of conditions:
\begin{align}
\Delta b_{3}^{LR} = \Delta b_{2}^{LR} & \equiv \Delta b,
\\\Delta b_{B-L}^{LR}+\frac{3}{2} \Delta b_{R}^{LR}-11 &= 
\frac{5}{2}(\Delta b).
\end{align}
Obviously, many different sets of $\Delta b^{'}s$ can fulfil these
conditions and also realize particle configurations that provide a
realistic CKM. To provide just the simplest example, consider scalar
CKM models, class [a.1]. These require at least one copy of
$\Phi_{1,1,3,0}$ and $\Phi_{1,1,3,-2}$ each, as discussed in the
previous subsection. The simplest sliding solution for this class is
given by $\Phi_{1,1,3,0}+4\Phi_{1,1,3,-2}$ with $\Delta
b^{'}s=(0,0,10/3,6)$ (and a $m_{G}=2 \times 10^{16}$ GeV). In the 
LR regime we thus have SM (+ Higgs completed to one bi-doublet) particle 
content plus $\Psi_{1,2,2,0}+\Psi_{1,3,1,0}+\Psi_{8,1,1,0}+\Phi_{1,1,3,0}
+4\Phi_{1,1,3,-2}$. Fig.~\ref{fig:slidingGCU} shows the independence 
of the GCU from the value of $m_{LR}$. Note again, that GCU is lost, 
once $m_{NP}$ is raised above a certain value, the b.f.p. for 
$m_{NP}$, including 2-loop coefficients, being $m_{NP}=1.1$ TeV. 

As in the case of the non-sliding solutions, of course it is also
possible to construct sliding-LR models of class [a.2], the simplest
2-field solution is $2\Phi_{1,1,2,-1}+20\Phi_{1,1,1,2}$ with $\Delta
b^{'}_i=(0,0,1/3,21/2)$.

As mentioned above, unification in non-SUSY extensions of the SM 
have been studied already in \cite{Amaldi:1991zx}. A particular 
interesting example is the one studied in \cite{Gogoladze:2010in}, 
which adds two kinds of VLQs to the SM particle content, namely 
$Q'=\Psi_{3,2,1/6}$ and $d^{'c}=\Psi_{{\bar 3},1,1/3}$. This model 
could, potentially, explain the much discussed enhancement in 
$h\to\gamma\gamma$ \cite{ATLAS:2013oma,Chatrchyan:2013lba}. 
\footnote{The latest CMS data now gives much smaller $h\to\gamma\gamma$, 
see the web-page of CMS public results at: 
twiki.cern.ch/twiki/bin/view/CMSPublic/PhysicsResultsHIG. }

\begin{figure}[htb]
\centering
\includegraphics[width=0.52\linewidth]{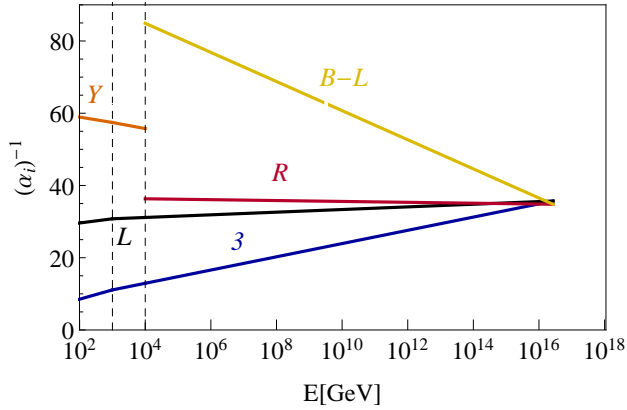}
\caption{Evolution of the inverse gauge couplings in the second example of 
sliding-LR models: $2\Psi_{3,2,1,1/3}+2\Psi_{3,1,1,2/3}+2\Phi_{1,1,1,2}
+4\Phi_{1,1,3,-2}$. This example is non-SUSY and with a CKM explained 
by VLQs (class [b]).}
\label{fig:sliding2}  
\end{figure}

As our second sliding-LR example, we thus choose 
$X=2\Psi_{3,2,1/6}+2\Psi_{3,1,1/3}+\Phi_{1,2,1/2}$, which in the 
LR regime corresponds to $X = 2\Psi_{3,2,1,1/3}+2\Psi_{3,1,1,2/3}$, with 
the $\Phi_{1,2,1/2}$ used to complete the $\Phi_{1,2,2,0}$. 
The $\Delta b^{LR}_{i}$ coefficients of this configuration are:
\begin{equation}
(\Delta b_{3}^{LR},\Delta b_{2}^{LR}, \Delta b_{R}^{LR}, \Delta b_{B-L}^{LR}) = 
(2,2,0,1).
\end{equation}
The sliding conditions in this case are the same as above and the 
simplest solution following these conditions and allowing to break 
the LR symmetry correctly is: $2\Phi_{1,1,1,2}+4\Phi_{1,1,3,-2}$, 
with $\Delta b^{'}_i=(0,0,8/3,7)$. The running of the inverse 
gauge couplings for this example is shown in fig.~\ref{fig:sliding2}.

\section{Uncertainties in new physics scale and proton half-life}
\label{sec:ChiSq}

One of the aspects of model building for new physics models, rarely
discussed in the literature, are uncertainties. While ideally, of
course, predictions such as the existence of new particles at the TeV
scale should be testable over the whole range of the allowed parameter
space, in reality most model builders content themselves with showing
that for some particular choice of parameters consistent solutions for
their favorite model exist.

In this section we discuss uncertainties for the predictions of our LR
models. In these models, once we have fixed the particle content of a
particular version, there are essentially three free parameters:
$m_{LR}$, $m_G$ and $\alpha_G$. However, since there are also three
gauge couplings, with values fixed by experiment, for any given model
$m_{LR}$, $m_G$ and $\alpha_G$ are fixed up to some error by the
requirement of gauge coupling unification. This results essentially in
two predictions: First, the mass scale, where the gauge bosons of the
extended gauge sector and (possibly) other particles of the model
should show up. This scale coincides, of course, with the range of
$m_{LR}$, as derived from the fit.  And, second, derived from $m_G$ and
$\alpha_G$, we obtain a range for the predicted half-life of proton decay.

The analysis of this section uses a $\chi^2$ minimization, which fits
the three measured SM gauge couplings as functions of the three
unknowns.  We start by discussing the error budget. The total error
budget can be divided into a well defined experimental error plus a
theory error. For the experimental input we use
\cite{Beringer:1900zz}:
\begin{eqnarray}\label{eq:alpexp}
\alpha_1^{-1} = 58.99 \pm 0.020   \\ \nonumber
\alpha_2^{-1} = 29.57 \pm  0.012 \\ \nonumber
\alpha_3^{-1} = 8.45  \pm  0.050\ . 
\end{eqnarray}
The experimental errors quoted are at the 1-$\sigma$ confidence level (CL).  
Note especially the small value of $\Delta(\alpha_3^{-1})$, 
according to \cite{Beringer:1900zz}, compared to the older 
value of $\Delta(\alpha_3^{-1}) \simeq 0.14$~\cite{Amsler:2008zzb}.

Much more difficult to estimate is the theory error. In our discussion
presented in section~(\ref{sec:LR}) we have used 1-loop
$\beta$-coefficients for simplicity. Two-loop $\beta$-coefficients for
general non-supersymmetric theories, have been derived long ago
\cite{Jones:1981we,Machacek:1983tz,Machacek:1983tz}, see also
\cite{Luo:2002ti}, and can be easily included in a numerical analysis.
However, a {\em consistent} 2-loop calculation requires the inclusion
of the 1-loop thresholds from both, light states at the LR-scale and
heavy states at the GUT scale. While we do fix in our constructions
the particle content in the LR-symmetric phase, we have not specified
the Higgs content for the breaking of $SO(10)$ to the LR group 
in detail. Thus, the calculation of the GUT scale thresholds is not 
possible for us, even in principle. The ignorance of the thresholds 
should therefore be included as (the dominant part of) the theoretical 
error, once two-loop $\beta$-coefficients are used in the calculation.

The 1-loop thresholds are formally of the order of a 2-loop effect and, thus,  
it seems a reasonable guess to estimate their size by a comparison 
of the results using 1-loop and 2-loop $\beta$ coefficients in the 
RGE running. This, however, can be done using different assumptions. 
We have tried the following four different definitions for the 
theory error:
\begin{itemize}
\item (i) Perform a $\chi^2_{min}$ search at 1-loop and at 2-loop.
Consider the difference $\Delta(\alpha_G^{-1})^{\rm th} \simeq
|(\alpha_G^{-1})^{\rm (1-loop)} - (\alpha_G^{-1})^{\rm(2-loop)}|$ 
as the theoretical error, common to all $\alpha_i$.
\item (ii) Perform a $\chi^2_{min}$ search at 1-loop and at 2-loop.
Calculate $\Delta(\alpha_i^{-1})^{\rm th} \simeq |(\alpha_i^{-1})^{\rm
exp} - (\alpha_i^{-1})^{\rm (2-loop)}|$ using $m_{G}^{\rm 1-loop}$ as the
starting point, but keeping the $m_{LR}$ and $\alpha_G^{-1}$ from the 
2-loop calculation. This generates $\Delta(\alpha_i^{-1})^{\rm th}$ which
depend on the group $i$, but does not take into account the overall
shift on $\alpha_G^{-1}$ caused by the change from 1-loop to 2-loop
coefficients.
\item (iii) Perform a $\chi^2_{min}$ search at 1-loop and at 2-loop.
Calculate $\Delta(\alpha_i^{-1})^{\rm th} \simeq |(\alpha_i^{-1})^{\rm
exp} - (\alpha_i^{-1})^{\rm (2-loop)}|$ using $m_{G}^{\rm 1-loop}$
{\em and} $(\alpha_G^{-1})^{\rm 1-loop}$ as the starting point, but
keeping the $m_{LR}$ from the 2-loop calculation. This takes into
account both, the shift of $m_G$ and $\alpha_G^{-1}$ from 1-loop to
2-loop calculation.
\item (iv) Perform a $\chi^2_{min}$ search at 1-loop. For the
b.f.p. of $m_G$, $\alpha_G^{-1}$ and $m_{LR}$ found, calculate the
values of $(\alpha_i^{-1})^{\rm (2-loop)}$. Use
$\Delta(\alpha_i^{-1})^{\rm th} \simeq |(\alpha_i^{-1})^{\rm exp} -
(\alpha_i^{-1})^{\rm (2-loop)}|$ as the error. One should expect this
definition to give, in principle, the most pessimistic error
estimate. See, however, the discussion below.
\end{itemize}

\begin{table}[t]
\begin{center}
\scalebox{0.95}{
\begin{tabular}{ | l | l | l | l |l |}
\hline
Def.: & $\Delta(\alpha_1^{-1})$ & $\Delta(\alpha_2^{-1})$ 
&$\Delta(\alpha_3^{-1})$ & $\overline{\Delta}(\alpha^{-1})$ \\
\hline
(i) & 0.76 & 0.76 & 0.76  & 0.76 \\
(ii) & 0.57 & 0.41 & 1.18 & 0.72 \\
(iii) & 1.31 & 0.34  & 0.40 & 0.68\\
(iv) & 1.21 & 0.41   & 0.40 & 0.67 \\
\hline
\end{tabular}
}
\scalebox{0.95}{
\begin{tabular}{ | l | l | l | l |l |}
\hline
Def.: & $\Delta(\alpha_1^{-1})$ & $\Delta(\alpha_2^{-1})$ 
&$\Delta(\alpha_3^{-1})$ & $\overline{\Delta}(\alpha^{-1})$\\
\hline
(i) & 0.86 & 0.86 & 0.86 & 0.86\\
(ii) & 0.46 & 0.46 & 1.18 & 0.70 \\
(iii) & 1.30 & 0.39 & 0.30 & 0.66 \\
(iv) & 1.11 & 0.44 & 0.22 & 0.59 \\
\hline
\end{tabular}
}
\end{center}
\caption{Example shifts (``errors'') in $\Delta(\alpha_i^{-1})$ for the
particular models: SM + $\Phi_{1,2,2,0}+3 \Phi_{1,1,3,0}
+2\Phi_{1,1,3,-2}$ (left) and SM + $2\Psi_{3,1,1,-2/3}+2 \Phi_{1,2,1,1}
+2\Phi_{1,1,3,-2}$ (right), see also fig.~\ref{fig:LRlow},
determined using the four different methods defined in the text. 
$\overline{\Delta}(\alpha^{-1})$ is the mean deviation.}
\label{tab:errors}
\end{table}

Example shifts (``errors'') in $\Delta(\alpha_i^{-1})$ determined by
the four different methods defined above and for two particular
models, discussed in previous sections, are given in 
table~\ref{tab:errors}. The first and most important observation is that
the theory errors estimated in this way are always much larger than
the experimental errors on the gauge couplings.  We would like to
stress, however, that in absolute terms $\Delta(\alpha_G^{-1})^{\rm
th} \simeq 0.5$ corresponds only to a $1\div 2$~\% shift in the value of
$\alpha_G^{-1}$, depending on the model.  It is found that all four
methods lead to very similar $\overline{\Delta}(\alpha^{-1})$, but
which of the couplings is assigned the smallest error depends on the
method and on the model. 

\begin{figure}[t]
\includegraphics[width=0.48\linewidth]{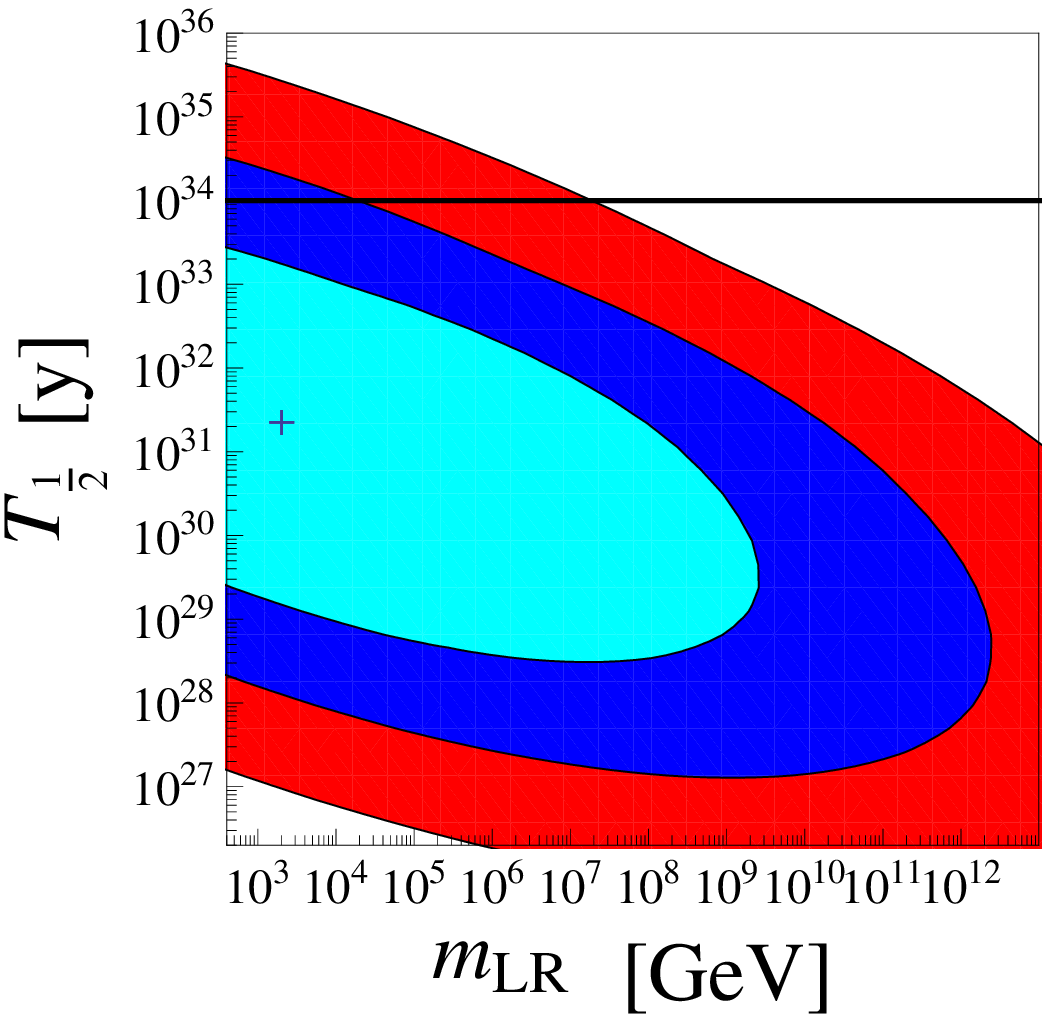}\hskip3mm
\hskip-2mm\includegraphics[width=0.48\linewidth]{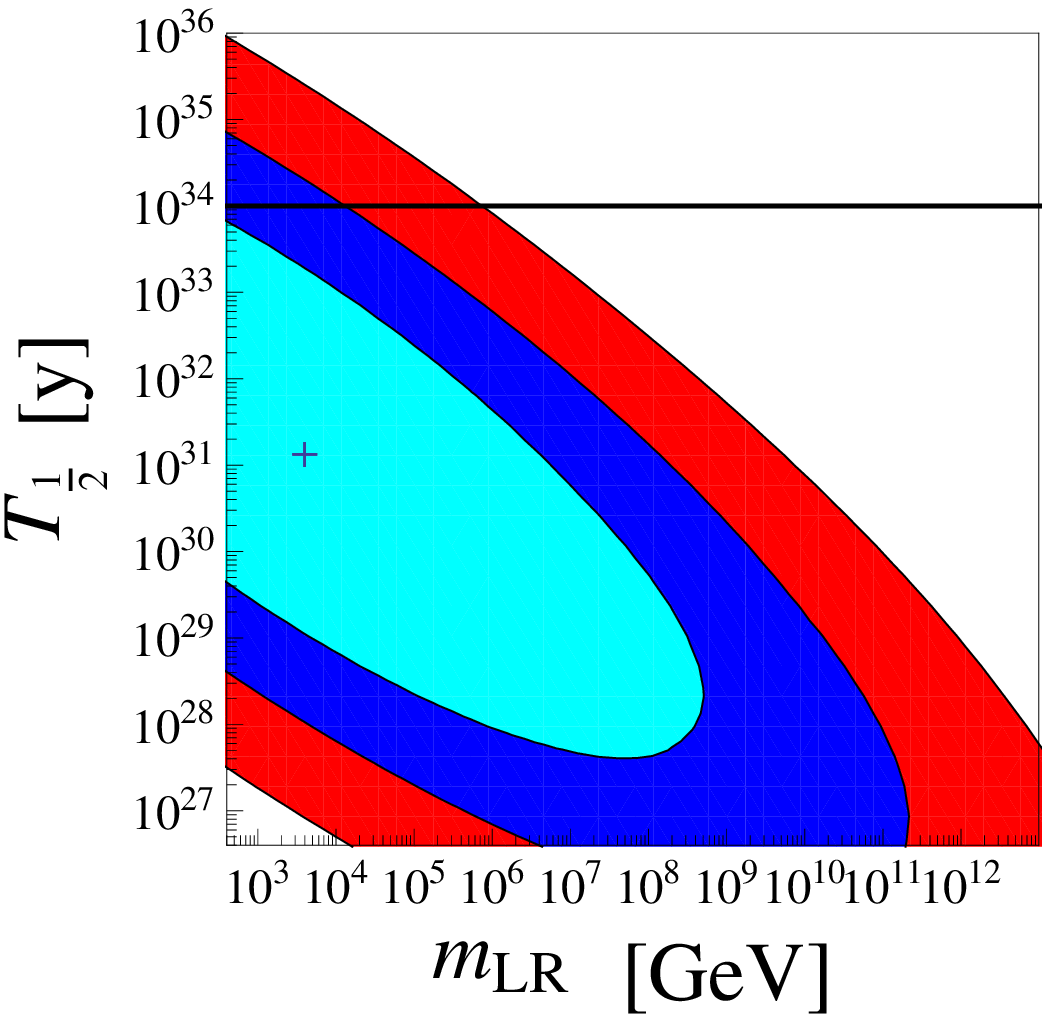}
\includegraphics[width=0.48\linewidth]{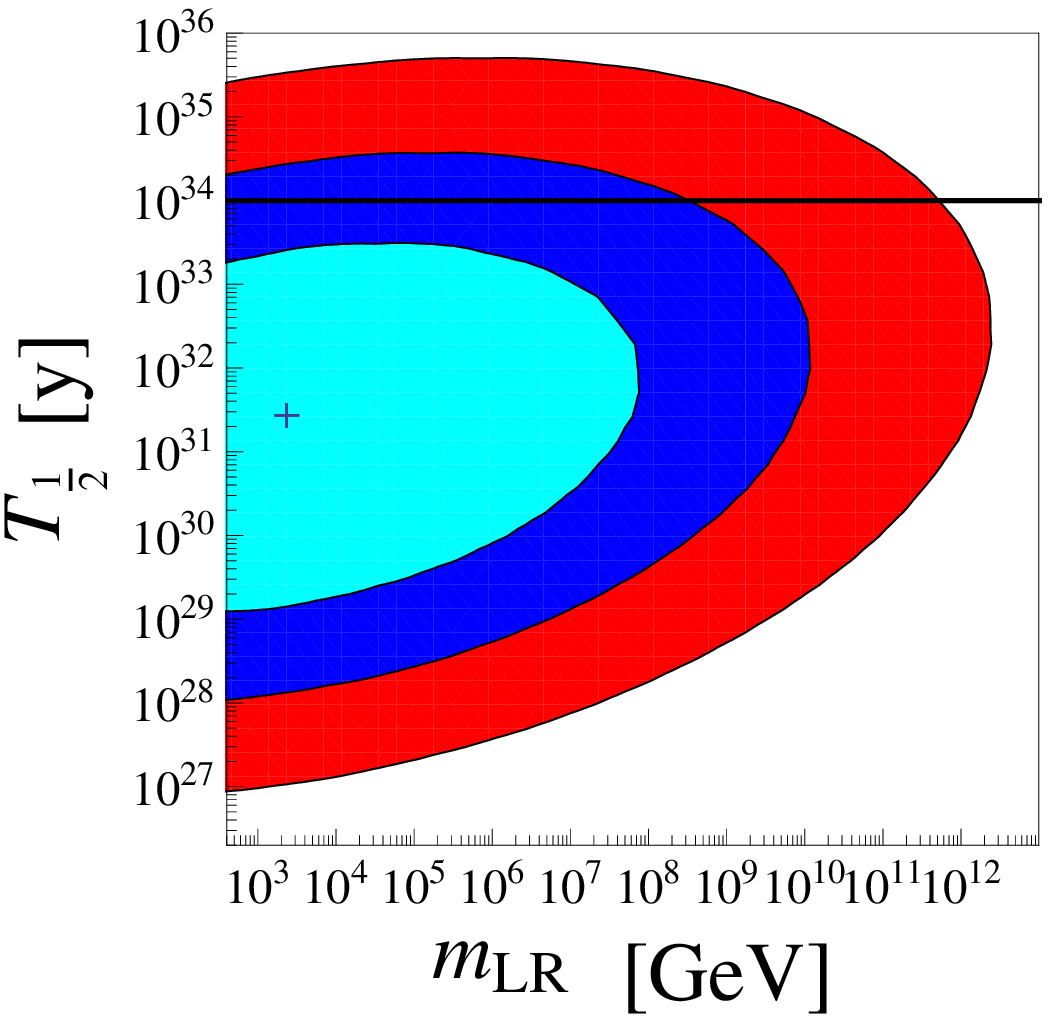}\hskip3mm
\hskip-2mm\includegraphics[width=0.48\linewidth]{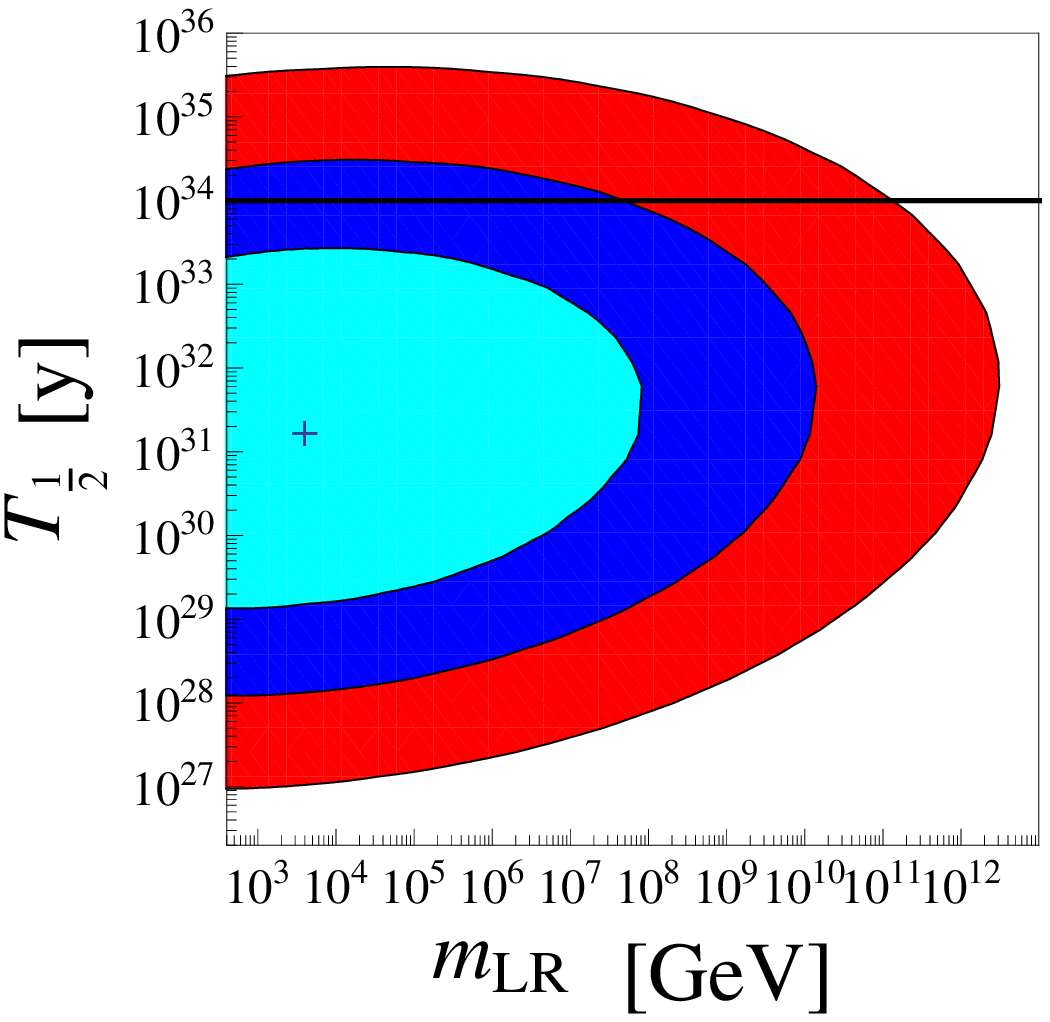}
\caption{Contour plot of the $\chi^2$ distribution in the plane
($m_{LR},T_{1/2}$) for the model: SM + $\Phi_{1,2,2,0}+3
\Phi_{1,1,3,0} +2\Phi_{1,1,3,-2}$, using the four different approaches
to estimate the theoretical error, defined in the text: Top row: (i)
left and (ii) right, bottom row (iii) left and (iv) right.  The cyan
(blue, red) region corresponds to the allowed region at 68 \% (95 \%
and 3-$\sigma$) CL. In all four cases the model is ruled out by proton 
decay constraints at one sigma, but allowed at 2-$\sigma$ CL. 
For further  discussion see text.
\label{fig:ChisqLRErr}}
\end{figure}  

Perhaps more surprising is that method (iv) in the examples shown in
the table {\em does not} automatically lead to the largest
$\Delta(\alpha_i^{-1})$ nor to the largest average error,
$\overline{\Delta}(\alpha^{-1})$, in these examples
\footnote{For the MSSM method (iv) indeed leads to the largest
$\Delta(\alpha_i^{-1})$, see below.}. We can attribute this somewhat 
unexpected result to the correlated shifts induced by the simultaneous 
change in $m_{LR}$ and $m_G$ in method (iv), which can even conspire 
in some models to give an unrealistically small deviation in one 
particular coupling, see the value of $\Delta(\alpha_3^{-1})$ in 
the second model shown in table~\ref{tab:errors}, for example. 

In fig.~\ref{fig:ChisqLRErr} we then show the $\chi^2$ distributions
using the four different set of values of $\Delta(\alpha_i^{-1})$ for
the model used in the left panel of table~\ref{tab:errors}. Here,
the $\chi^2_{\rm min}$ (denoted by the cross) and the corresponding 1,
2- and 3-$\sigma$ CL contours are shown in the plane
$(m_{LR},T_{1/2})$, where $T_{1/2}$ is the proton decay half-life
estimated via eq.~(\ref{eq:ProtonLifetime}).  While at first glance,
the different methods seem to produce somewhat different results, a
closer inspection reveals that the two main conclusions derived from
this analysis are in fact independent of the method. First, in all
four methods the model is excluded by the lower limit for the proton
decay half-live from Super-K \cite{Nishino:2012ipa,Abe:2013lua} data
at the one sigma level, but becomes (barely) allowed at 2-$\sigma$
CL.  And, second, while the model has a preferred value for the
$m_{LR}$ scale within the reach of the LHC, the upper limit on
$m_{LR}$ - even at only 1-$\sigma$ CL! - is very large, between
[$5\times 10^{7},2\times 10^{9}$] GeV depending on the method. The
model could therefore be excluded by (a) a slight improvement in the
theoretical error bar or (b) from an improved limit on the proton
decay, but not by direct accelerator searches. This latter conclusion
is, of course, not completely unexpected, since the value of $m_{LR}$
enters in the analysis only logarithmically as the difference between
$m_G$ and $m_{LR}$.

As fig.~\ref{fig:ChisqLRErr} shows, in three of the four methods the
error in the determination of the $T_{1/2}$ is around $2\div 2.5$ orders
of magnitude at one sigma, while in method (ii) - due to the
correlation with $m_{LR}$ - we find approximately $T_{1/2} = 
10^{31+2.8-3.5}$ y. This is mainly due to a change in the GUT scale,
when going from the 1-loop to the 2-loop $\beta$-coefficients. Note,
that the value of $m_G$ enters in the fourth power in the calculation
of $T_{1/2}$; thus, an error of a factor of 100 corresponds only to a
shift of a factor of $\Delta (m_G) \simeq 3$ in the GUT scale.

\begin{figure}[t]

\includegraphics[width=0.48\linewidth]{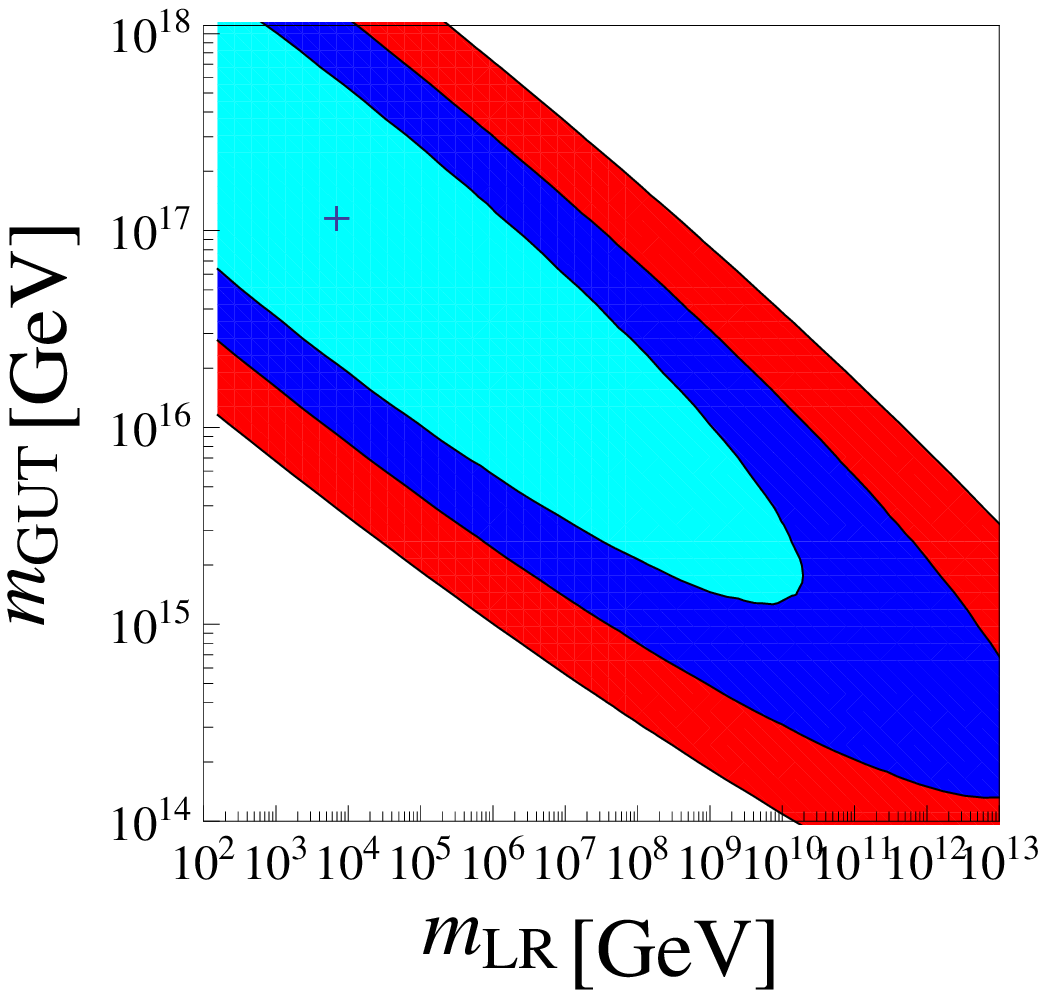}\hskip3mm
\hskip-2mm\includegraphics[width=0.48\linewidth]{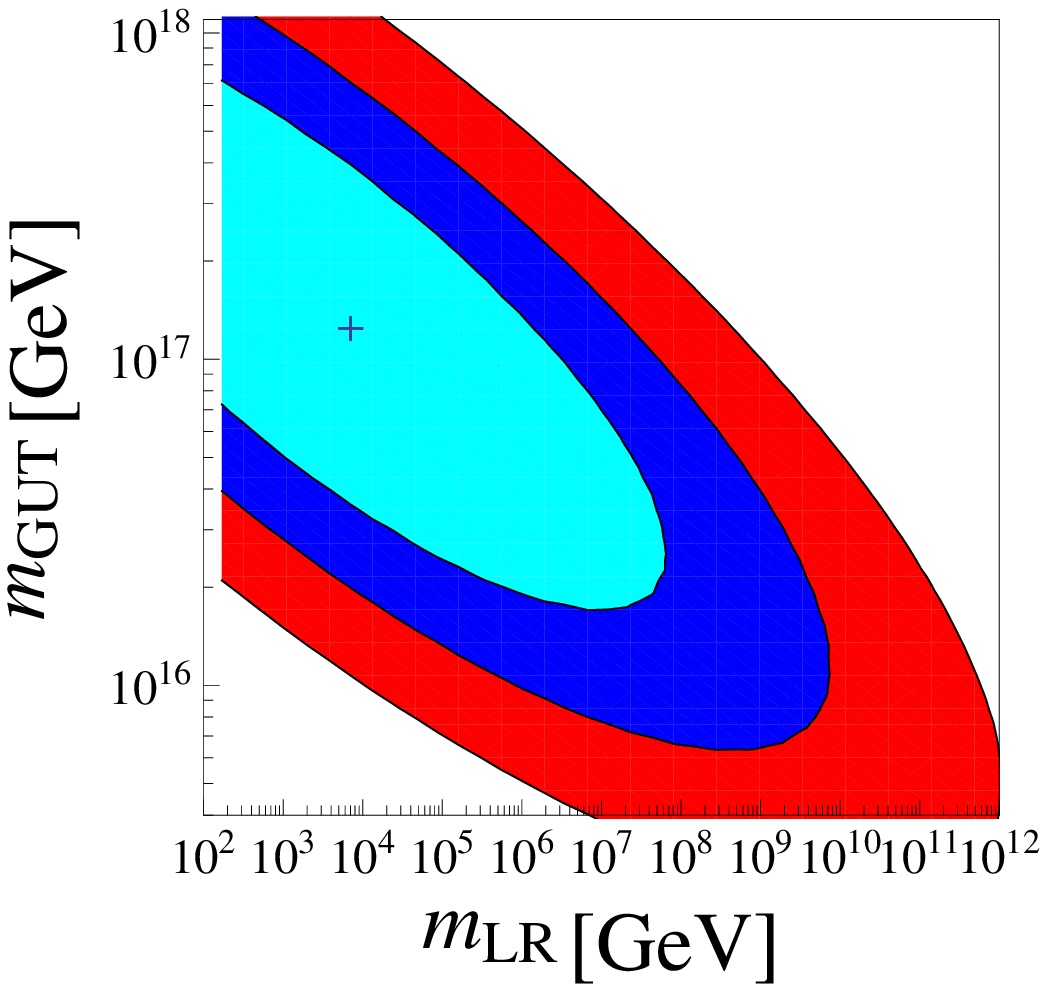}
\caption{Contour plot of the $\chi^2$ distribution in the plane
  ($m_{LR},m_{G}$) for the model: SM + $2\Psi_{3,1,1,-2/3}+2
  \Phi_{1,2,1,1} +2\Phi_{1,1,3,-2}$, using the four different
  approaches to estimate the theoretical error, defined in the text:
  (i) left and (iv) right. Methods (ii) and (iii) lead to results
  similar to (i) and (iv), respectively, and are therefore not shown.
\label{fig:ChisqLRErr2}}
\end{figure}  

In fig.~\ref{fig:ChisqLRErr2} we show the $\chi^2$ distributions,
for the model on the right panel of table~\ref{tab:errors}, using
two of the four methods for determining $\Delta(\alpha_i^{-1})$ of
table~\ref{tab:errors}. The plots for methods (ii) and (iii) lead to
results similar to (i) and (iv), respectively, and are therefore not
shown. Again, $m_{LR}$ is only very weakly constrained in this
analysis, but for this model, the b.f.p. of the GUT scale is much
larger, around $m_G \simeq 10^{17}$ GeV, so proton decay provides
hardly any constraints on this model. Note the strong correlation
between $m_{LR}$ and $m_G$ in the plot on the left, which leads to a
much larger ``error'' bar in the predicted range of the proton decay
half-life for this model, roughly 5 orders of magnitude at one sigma
CL.

We have repeated this exercise for a number of different LR models 
\footnote{Among them the two ``minimal'' LR models discussed in the
  introduction.}, see the appendix and discussion in the previous
section and have always found numbers of similar magnitude. We have
checked, however, that these ``large'' shifts in
$\Delta(\alpha_i^{-1})$ {\em are not a particular feature of our LR
models}. For this check we have calculated $\Delta(\alpha_G^{-1})^{\rm
  th}$ also for a number of models with only the SM group up to the
GUT scale (see appendix).  There, instead of $m_{LR}$ we used the
energy where the new particles appear, call it $m_{NP}$, as a free
parameter. Very similar values and variations for
$\Delta(\alpha_i^{-1})^{\rm th}$ are found in this study too. It may
be interesting to note that the smallest $\Delta(\alpha_G^{-1})^{\rm
  th}$ we found corresponds to a model which is essentially like split
supersymmetry \footnote{This is the first example of SM+X
  configurations discussed in section~\ref{subsect:SlidLR}.}  with a
$\Delta(\alpha_G^{-1})^{\rm th}$ of only $\Delta(\alpha_G^{-1})^{\rm
  th} \simeq 0.25$. (In methods (ii)-(iv) the
$\Delta(\alpha_G^{-1})^{\rm th}$ vary for this model between $0.05$
and $0.78$ with a mean of $0.55$.)  On the other hand, for the MSSM we
find a $\Delta(\alpha_G^{-1})^{\rm th} \simeq 0.82$ and values of
$\Delta(\alpha_i^{-1})^{\rm th}$ even up to $\Delta(\alpha_i^{-1})^{\rm th}
\simeq 2$, depending on which of our four methods is used. Thus, the
uncertainties discussed in this section should apply to practically
all new physics models, which attempt to achieve GCU.

\begin{figure}[t]
\includegraphics[width=0.48\linewidth]{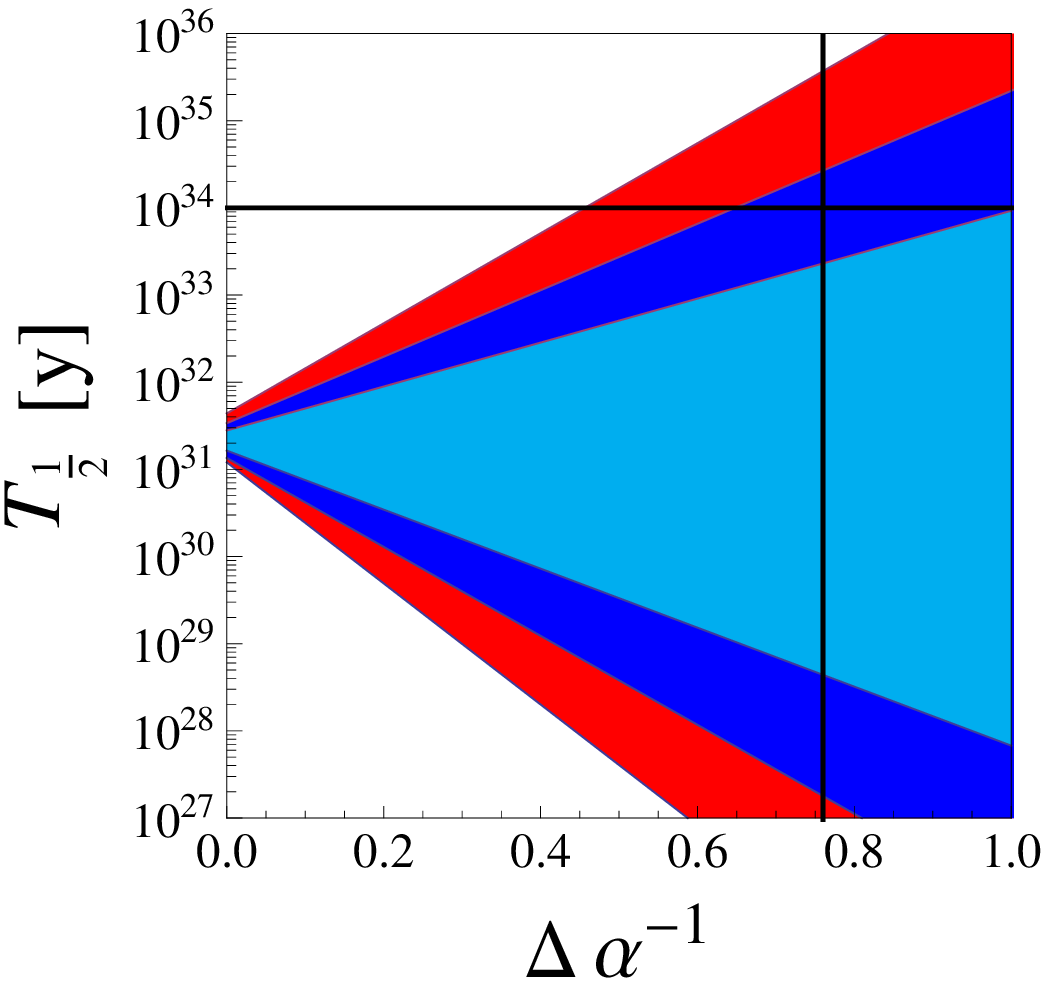}\hskip3mm
\hskip-2mm\includegraphics[width=0.48\linewidth]{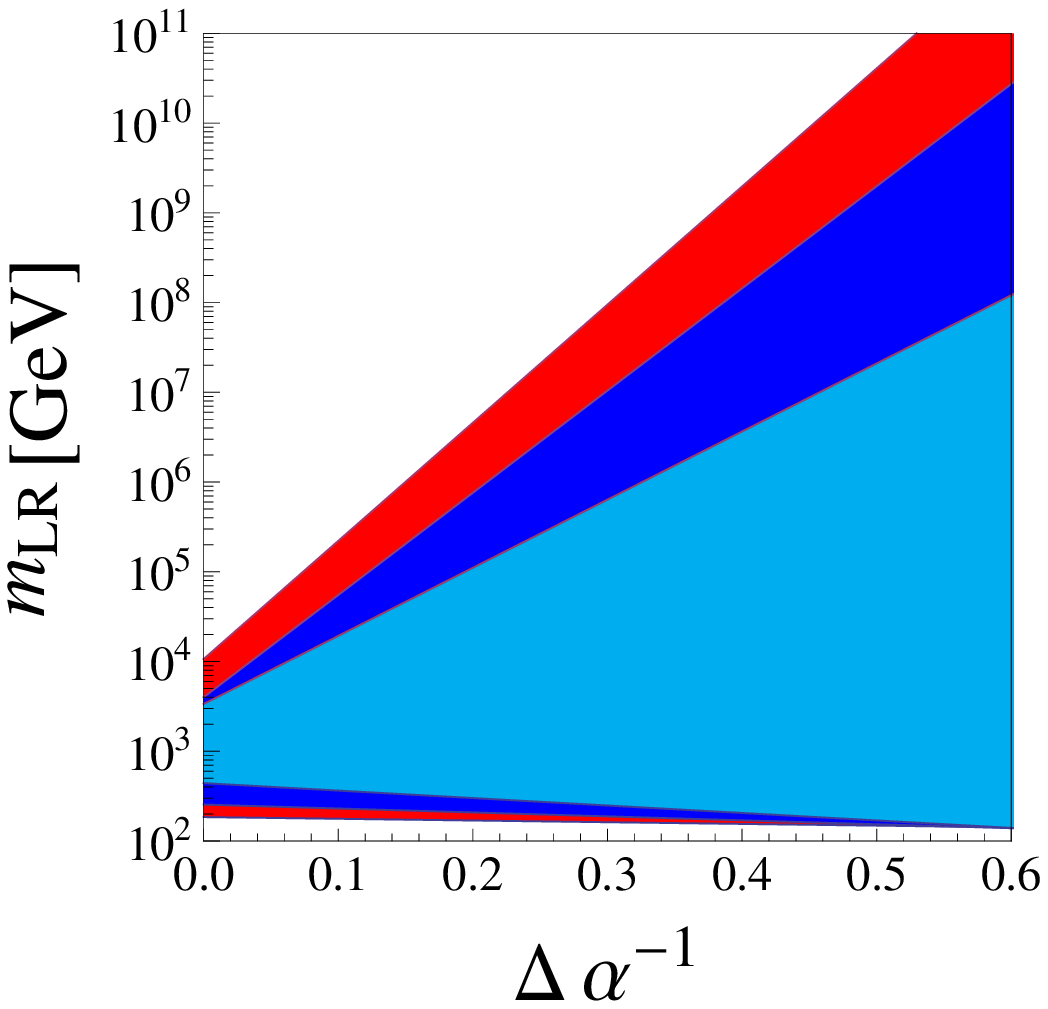}
\caption{Allowed range  cyan (blue, red)  of $T_{1/2}$ (left) 
and $m_{LR}$ (right) 
at 1-, 2- and 3-$\sigma$ CL as a function of the error in 
$\Delta(\alpha^{-1})$. 
The plot is for the model SM + $\Phi_{1,2,2,0}+3
\Phi_{1,1,3,0} +2\Phi_{1,1,3,-2}$. The horizontal line in the left 
plot is the experimental lower limit \cite{Nishino:2012ipa,Abe:2013lua}, 
while the vertical line at $\Delta(\alpha^{-1})=0.76$ corresponds to 
the estimated uncertainty in this model using method (i).}

\label{fig:ChisqtP2}
\end{figure}  

Reducing the theory error on $\alpha_i^{-1}$ will be possible only, if
thresholds are calculated at {\em both new physics scales, $m_{LR}$
  and $m_G$.} Since this task is beyond the scope of the present work,
in fig.~\ref{fig:ChisqtP2} we show plots as a function of the
unknown theory error $\Delta(\alpha^{-1})$. The model considered is
excluded by the proton decay constraint at 2-$\sigma$ CL up to an error 
of roughly $\Delta(\alpha^{-1})\simeq 0.6$, 
indicating that even a minor improvement in the theory
error can have important consequences for all models with a relatively
low GUT scale, say $m_G \sim (1-3) \times 10^{15}$ GeV.  On the other
hand, in order to be able to fix the LR-scale to a value low enough
such that accelerator tests are possible, requires a much smaller
theory error. The exact value of this ``minimal'' error required
depends on the model, but as can be seen from fig.~\ref{fig:ChisqtP2}
theory errors of the order of 
 $\Delta(\alpha_i^{-1}) \lsim 0.1$ will be necessary.

\section{Conclusions}
\label{sec:conclusions}

In this work we attempted to construct a comprehensive list of
non-SUSY models with LR-symmetric intermediate stage close to the TeV
scale that may be obtained as simple low-energy effective theories
within a class of renormalizable non-SUSY SO(10) grand unifications
assuming some of the components of scalar representations with
dimensions up to 126 to be accidentally light. In order to make our
way through the myriads of options we assumed that all such light
fields (besides those pushed down by the need to arrange for the low
LR breaking scale) necessary to maintain the SO(10)-like gauge
coupling unification are clustered around the same (TeV)
scale. 

Remarkably enough, the vast number of settings that pass all the
phenomenological constraints (in particular, the compatibility with
the quark and lepton masses and mixings, the current proton lifetime
limits, perturbativity and gauge coupling unification) can be grouped
into a relatively small number of types characterised, in our
classification, by the extra fields underpinning the emergence of the
SM flavour structure.  Needless to say, the popular low-scale LR
alternatives to the MSSM such as, e.g., split-SUSY, simple extensions
of the mLR and/or m$\Omega$LR models, are all among these. 

In the second part of the study we elaborate in detail on the
theoretical uncertainties affecting the possible determination of (not
only) the LR scale from the low-energy observables focusing namely on
the impact of different definitions of the $\chi^{2}$ reflecting the
generic incapability of the simplistic bottom-up approach to account
for most of the details of the full top-down analysis. To this end, we
perform a numerical analysis of a small set of sample scenarios to
demonstrate how difficult it is in general to extrapolate the
low-energy information over the ``desert'' to draw any strong
conclusion about the viability of the underlying unified theory
without a detailed account for, e.g., the GUT-scale thresholds and
other such high-scale effects. Nevertheless, within the bottom-up
approach employed in this study the character of our results is
inevitably just indicative and further improvements are necessary
before drawing any far-fetched conclusions. To this end, the simple
classification of the basic potentially realistic schemes given in
Sect.~\ref{sec:LR} may be further improved in several directions,
among which perhaps the most straightforward are, e.g., the viability
of arranging the considered spectra in specific SO(10) GUTs, their
perturbativity beyond the unification scale, etc.

\section*{Acknowledgements}

This work is supported in part by EU~Network grant UNILHC
PITN-GA-2009-237920. C.A. and M.H. also acknowledge support from the
Spanish MICINN grants FPA2011-22975, MULTIDARK CSD2009-00064 and the
Generalitat Valenciana grant Prometeo/2009/091. J. C. R. acknowledges
the financial support from grants CFTP-FCT UNIT 777,
CERN/FP/123580/2011 and PTDC/FIS/102120/2008. The work of M.M. is
supported by the Marie-Curie Career Integration Grant within the 7th
European Community Framework Programme FP7-PEOPLE-2011-CIG, contract
number PCIG10-GA-2011-303565, by the Research proposal
MSM0021620859 of the Ministry of Education, Youth and Sports of the
Czech Republic and by the ``Neuron'' Foundation for scientific research.

\appendix
\section{List of fields}
\label{sec:appendix}

Table (\ref{tab:List_of_LR_fields}) gives transformation properties 
under the group $SU(3)_{c}\times SU(2)_{L} \times SU(2)_{R} \times U(1)_{B-L}$ 
for all representations of $SO(10)$ up to dimension $126$. For the 
sake of convenience only, we also give names of certain representations, 
which have been used in the literature before.


\begin{table}[t]
\begin{center}
\scalebox{0.98}{

\begin{tabular}{c}
\begin{tabular}{|l|rrrrrrrrrrrrrr|}
\toprule 
 & $1$ & $2$ & $3$ & $4$ & $5$ & $6$ & $7$ & $8$ & $9$ & $10$ & $11$ & $12$ & $13$ & $14$ \tabularnewline
\midrule 
 Scalar&  &$\chi$ & $\chi^{c}$ & $\Omega$ & $\Omega^{c}$ & $\Phi$ & $ $ & $ $ & $ $ & $ $ & $ $ & $ $ & $ $ & $ $ \tabularnewline
\midrule
Fermion& $\widetilde{B}$ &$L$ & $L^{c}$ &$\Sigma$ & $\Sigma^{c}$ &  & $\widetilde{G}  $ & $ $ & $\delta_{d}$ & $\delta_{u}$ & $ $ & $  $ & $Q$ & $Q^{c} $\tabularnewline
\midrule
$SU(3)_{C}$ & \textbf{1} & \textbf{1} & \textbf{1} & \textbf{1} & \textbf{1} & \textbf{1} & \textbf{8} & \textbf{1} & \textbf{3} & \textbf{3} & \textbf{6} & \textbf{6} & \textbf{3} & \textbf{3}\tabularnewline
$SU(2)_{L}$ & \textbf{1} & \textbf{2} & \textbf{1} & \textbf{3} & \textbf{1} & \textbf{2} & \textbf{1} & \textbf{1} & \textbf{1} & \textbf{1} & \textbf{1} & \textbf{1} & \textbf{2} & \textbf{1}\tabularnewline
$SU(2)_{R}$ & \textbf{1} & \textbf{1} & \textbf{2} & \textbf{1} & \textbf{3} & \textbf{2} & \textbf{1} & \textbf{1} & \textbf{1} & \textbf{1} & \textbf{1} & \textbf{1} & \textbf{1} & \textbf{2}\tabularnewline
$U(1)_{B-L}$ & 0 & +1 & -1 & 0 & 0 & 0 & 0 & +2 & $-\frac{2}{3}$ & $+\frac{4}{3}$ & $+\frac{2}{3}$ & $-\frac{4}{3}$ & $+\frac{1}{3}$ & $+\frac{1}{3}$\tabularnewline
\midrule
\textbf{\footnotesize }%
\begin{tabular}{{@{}l@{}}}
\textbf{SO(10)}\tabularnewline
\bf{Origin}\tabularnewline
\end{tabular} & \textbf{\footnotesize }%
\begin{tabular}{{@{}r@{}}}
\textbf{\footnotesize $1$}\tabularnewline
\textbf{\footnotesize $54$}\tabularnewline
\textbf{\footnotesize $45$}\tabularnewline
\end{tabular} & \textbf{\footnotesize $16$} & \textbf{\footnotesize $\overline{16}$} & \textbf{\footnotesize $45$} & \textbf{\footnotesize $45$} & \textbf{\footnotesize }%
\begin{tabular}{{@{}r@{}}}
\textbf{\footnotesize $10$}\tabularnewline
\textbf{\footnotesize $120$}\tabularnewline
\textbf{\footnotesize $126$}\tabularnewline
\end{tabular} & \textbf{\footnotesize }%
\begin{tabular}{{@{}r@{}}}
\textbf{\footnotesize $45$}\tabularnewline
\textbf{\footnotesize $54$}\tabularnewline
\end{tabular} & \textbf{\footnotesize $120$} & \textbf{\footnotesize }%
\begin{tabular}{{@{}r@{}}}
\textbf{\footnotesize $10$}\tabularnewline
\textbf{\footnotesize $126$}\tabularnewline
\textbf{\footnotesize $120$}\tabularnewline
\end{tabular} & \textbf{\footnotesize $45$} & \textbf{\footnotesize $120$} & \textbf{\footnotesize $54$} & \textbf{\footnotesize $16$} & \textbf{\footnotesize $\overline{16}$}\tabularnewline
\bottomrule
\end{tabular}\tabularnewline
\tabularnewline
\begin{tabular}{|l|rrrrrrrrrr|}
\toprule 
 & $15$ & $16$ & $17$ & $18$ & $19$ & $20$ & $21$ & $22$ & $23$ & $24$\tabularnewline
\midrule
 Scalar& & $\Delta$ & $\Delta^{c}$ &  &  &  &  &  &  & \tabularnewline
\midrule
 Fermion& & &  &  &  &  &  &  &  & \tabularnewline
\midrule
$SU(3)_{C}$ & \textbf{8} & \textbf{1} & \textbf{1} & \textbf{3} & \textbf{3} & \textbf{3} & \textbf{6} & \textbf{6} & \textbf{1} & \textbf{3}\tabularnewline
$SU(2)_{L}$ & \textbf{2} & \textbf{3} & \textbf{1} & \textbf{2} & \textbf{3} & \textbf{1} & \textbf{3} & \textbf{1} & \textbf{3} & \textbf{2}\tabularnewline
$SU(2)_{R}$ & \textbf{2} & \textbf{1} & \textbf{3} & \textbf{2} & \textbf{1} & \textbf{3} & \textbf{1} & \textbf{3} & \textbf{3} & \textbf{2}\tabularnewline
$U(1)_{B-L}$ & 0 & -2 & -2 & $+\frac{4}{3}$ & $-\frac{2}{3}$ & $-\frac{2}{3}$ & $+\frac{2}{3}$ & $+\frac{2}{3}$ & 0 & $-\frac{2}{3}$\tabularnewline
\midrule
\textbf{\footnotesize }%
\begin{tabular}{{@{}l@{}}}
\textbf{SO(10)}\tabularnewline
\bf{Origin}\tabularnewline
\end{tabular} & \textbf{\footnotesize $120$} & \textbf{\footnotesize $126$} & \textbf{\footnotesize $\overline{126}$} & \textbf{\footnotesize } %
\begin{tabular}{{@{}r@{}}}
\textbf{\footnotesize $120$}\tabularnewline 
\textbf{\footnotesize $126$}\tabularnewline
\end{tabular} & \textbf{\footnotesize }%
\begin{tabular}{{@{}r@{}}}
\textbf{\footnotesize $120$}\tabularnewline
\textbf{\footnotesize $126$}\tabularnewline
\end{tabular} & \textbf{\footnotesize }%
\begin{tabular}{{@{}r@{}}}
\textbf{\footnotesize $120$}\tabularnewline
\textbf{\footnotesize $\overline{126}$}\tabularnewline
\end{tabular} & 
\textbf{\footnotesize $126$} & 
\textbf{\footnotesize $\overline{126}$} & 
\textbf{\footnotesize $54$} &   \textbf{\footnotesize } %
\begin{tabular}{{@{}r@{}}}
\textbf{\footnotesize $45$}\tabularnewline
\textbf{\footnotesize $54$}\tabularnewline
\end{tabular} 
\tabularnewline
\bottomrule
\end{tabular}
\end{tabular}

}
\end{center}

\caption{\label{tab:List_of_LR_fields}Naming conventions and
  transformation properties of fields in the left-right symmetric
  regime (not considering conjugates). The charges under the
  $U(1)_{B-L}$ group shown here were multiplied by a factor
  $\sqrt{\frac{8}{3}}$. The hypercharge is defined by:
  $Y=T^{R}_{3}+\frac{(B-L)}{2}$. $\widetilde{B}$ and $\widetilde{G}$
  correspond to the bino and gluino respectively. Symbols in the lines
  called "Scalar" and "Fermion" quote names used for these fields in the
  literature.}
\end{table}

\section{LR unification: simple configurations}

The breaking of the LR symmetry to the $SM$: $LR \rightarrow SM$
requires the presence of one of the fields: $\Phi_{1,1,3,-2}$ or
$\Phi_{1,1,2,-1}$. All configurations contain then at least one of
these fields and also one bi-doublet $\Phi_{1,2,2,0}$ (to complete
``$SM+$ bi-doublet'' basic field content). Table~\ref{tab:a.1} shows
the simplest LR configurations for $[a.1]$ scalar CKM (where the
necessary fields are: $\Phi_{1,1,3,0}$, $\Phi_{1,2,2,0}$,
$\Phi_{1,1,3,-2}$ or $\Phi_{1,1,2,-1}$) for each one of the fields
presented in table~\ref{tab:List_of_LR_fields}.

\begin{table}[t]
\begin{center}
\scalebox{0.95}{
\begin{tabular}{| l | l | l |  l | l |}
\hline
Extra field & Configuration & $\Delta b^{'}s$ & $m_{G}$ & $T_{1/2}$ \\ \hline
                & $\Phi_{1,2,2,0}+3\Phi_{1,1,3,0}+2\Phi_{1,1,3,-2}$ & $(0,\frac{1}{3},\frac{11}{3},3)$ & $2 \times 10^{15}$ & $10^{33 \pm 2.5}$ \\ \hline 

                & $\Phi_{1,2,2,0}+\Phi_{1,1,3,0}+3\Phi_{1,1,3,-2}$ & $(0,\frac{1}{3},3,\frac{9}{2})$ & $2 \times 10^{15}$ & $10^{33 \pm 2.5}$  \\ \hline 
               
$\Phi_{1,2,1,1}$  & $2\Phi_{1,2,2,0}+2\Phi_{1,1,3,0}+4\Phi_{3,1,1,-2/3}+\Phi_{1,2,1,1}+2\Phi_{1,1,3,-2}$ & $(\frac{2}{3},\frac{5}{6},\frac{10}{3},\frac{47}{12})$ & $8 \times 10^{15}$  & $10^{35\pm 2.5}$ \\ \hline
               
$\Phi_{1,1,2,-1}$ & $\Phi_{1,2,2,0}+2\Phi_{1,1,3,0}+2\Phi_{1,1,2,-1}+2\Phi_{1,1,3,-2}$ & $(0,\frac{1}{3},\frac{10}{3},\frac{7}{2})$ & $2 \times 10^{15} $ & $10^{33\pm 2.5} $\\ \hline

$\Phi_{1,3,1,0}$ & $\Phi_{1,2,2,0}+\Phi_{1,1,3,0}+\Phi_{1,3,1,0}+3\Phi_{1,1,3,-2}$ & $(1,1,3,\frac{9}{2})$ & $3 \times 10^{16} $ & $10^{37\pm 2.5} $\\ \hline
                                            
$\Phi_{8,1,1,0}$ &  $2\Phi_{1,2,2,0}+\Phi_{1,1,3,0}+\Phi_{8,1,1,0}+2\Phi_{1,1,3,-2}$ & $(1,\frac{2}{3},\frac{8}{3},3)$ & $4 \times 10^{17}$ & $10^{42\pm 2.5}$ \\ \hline
             
$\Phi_{1,1,1,2}$ & $\Phi_{1,2,2,0}+2\Phi_{1,1,3,0}+2\Phi_{1,1,1,2}+2\Phi_{1,1,3,-2}$ & $(0,\frac{1}{3},3,4)$ & $2 \times 10^{15}$  & $10^{33\pm 2.5}$ \\ \hline 
              
$\Phi_{3,1,1,-2/3}$ & $2\Phi_{1,2,2,0}+\Phi_{1,1,3,0}+5\Phi_{3,1,1,-2/3}+2\Phi_{1,1,3,-2}$ & $(\frac{5}{6},\frac{2}{3},\frac{8}{3},\frac{23}{6})$ & $1 \times 10^{17}$ & $10^{39\pm 2.5}$\\ \hline 
                
$\Phi_{3,1,1,4/3}$ &  $3\Phi_{1,2,2,0}+\Phi_{1,1,3,0}+4\Phi_{3,1,1,4/3}+2\Phi_{1,1,3,-2}$ & $(\frac{2}{3},1,3,\frac{17}{3})$ & $2\times 10^{15}$ & $10^{33\pm 2.5}$ \\ \hline 
                           
$\Phi_{6,1,1,2/3}$ &  $2\Phi_{1,2,2,0}+\Phi_{1,1,3,0}+\Phi_{6,1,1,2/3}+2\Phi_{1,1,3,-2}$ & $(\frac{5}{6},\frac{2}{3},\frac{8}{3},\frac{10}{3})$ & $1 \times 10^{17}$  & $10^{39\pm 2.5}$ \\ \hline 
                   
$\Phi_{6,1,1,-4/3}$ &  $2\Phi_{1,2,2,0}+3\Phi_{1,1,3,0}+\Phi_{6,1,1,-4/3}+\Phi_{1,1,3,-2}$ & $(\frac{5}{6},\frac{2}{3},\frac{10}{3},\frac{17}{6})$ & $1 \times 10^{17}$  &  $10^{39\pm 2.5}$\\ \hline 
                     
$\Phi_{3,2,1,1/3}$ &  $\Phi_{1,2,2,0}+\Phi_{1,1,3,0}+2\Phi_{3,1,1,-2/3}+\Phi_{3,2,1,1/3}+3\Phi_{1,1,3,-2}$ & $(\frac{2}{3},\frac{5}{6},3,\frac{59}{12})$ & $8 \times 10^{15}$ &  $10^{35\pm 2.5}$\\\hline  
$\Phi_{3,1,2,1/3}$ &  $\Phi_{1,2,2,0}+\Phi_{1,1,3,0}+\Phi_{3,1,2,/3}+2\Phi_{1,1,3,-2}$ & $(\frac{1}{3},\frac{1}{3},\frac{17}{6},\frac{37}{12})$ & $3 \times 10^{16}$  &  $10^{37\pm 2.5}$\\ \hline 
              
$\Phi_{8,2,2,0}$ &  $4\Phi_{1,2,2,0}+3\Phi_{1,1,3,0}+\Phi_{8,2,2,0}+3\Phi_{1,1,3,-2}$ & $(4,4,8,\frac{9}{2})$ & $3 \times 10^{16}$  &  $10^{37 \pm 2.5}$  \\\hline   
 
$\Phi_{1,3,1,-2}$ &  $\Phi_{1,2,2,0}+\Phi_{1,1,3,0}+2\Phi_{8,1,1,0}+2\Phi_{1,3,1-2}+2\Phi_{1,1,3,-2}$ & $(2,\frac{5}{3},\frac{7}{3},6)$ & $4 \times 10^{17}$ &  $10^{42 \pm 2.5}$  \\ \hline
              
$\Phi_{3,2,2,4/3}$ &  $\Phi_{1,2,2,0}+\Phi_{1,1,3,0}+\Phi_{8,1,1,0}+2\Phi_{3,2,2,4/3}+\Phi_{1,1,3-2}$ & $(\frac{7}{3},\frac{7}{3},4,\frac{11}{3}, \frac{41}{6})$ & $3 \times 10^{16}$ &  $10^{37 \pm 2.5}$  \\ \hline 

$\Phi_{3,3,1,-2/3}$ &  $\Phi_{1,2,2,0}+\Phi_{1,1,3,0}+2\Phi_{8,1,1,0}+\Phi_{3,3,1,-2/3}+4\Phi_{1,1,3-2}$ & $(\frac{5}{2},\frac{7}{3},\frac{11}{3},\frac{13}{2})$ & $1 \times 10^{17}$ &  $10^{37 \pm 2.5}$ \\ \hline 

$\Phi_{3,1,3,-2/3}$ &  $2\Phi_{1,2,2,0}+2\Phi_{1,1,3,0}+\Phi_{3,1,3,-2/3}+\Phi_{1,1,3,-2}$ & $(\frac{1}{2},\frac{2}{3},\frac{14}{3},2)$ & $8 \times 10^{15}$ &  $10^{35 \pm 2.5}$  \\ \hline

$\Phi_{6,3,1,2/3}$ &  $\Phi_{1,2,2,0}+3\Phi_{1,1,3,0}+2\Phi_{8,1,1,0}+\Phi_{6,3,1,2/3}+5\Phi_{1,1,3,-2}$ & $(\frac{9}{2},\frac{13}{3},\frac{17}{3},\frac{17}{2})$ & $1 \times 10^{17}$ &  $10^{39 \pm 2.5}$ \\ \hline  

$\Phi_{6,1,3,2/3}$ &  $2\Phi_{1,2,2,0}+\Phi_{1,1,3,0}+3\Phi_{1,3,1,0}+\Phi_{6,1,3,2/3}+2\Phi_{1,1,3,-2}$ & $(\frac{5}{2},\frac{8}{3},\frac{20}{3},4)$ & $8 \times 10^{15}$ &  $10^{35 \pm 2.5}$  \\ \hline  

$\Phi_{1,3,3,0}$ &  $2\Phi_{1,2,2,0}+3\Phi_{1,1,3,0}+3\Phi_{8,1,1,0}+\Phi_{1,3,3,0}+2\Phi_{1,1,3,-2}$ & $(3,\frac{8}{3},6,3)$ & $4 \times 10^{17}$ &  $10^{42 \pm 2.5}$ \\ \hline  

$\Phi_{3,2,2,-2/3}$ &  $\Phi_{1,2,2,0}+1\Phi_{1,1,3,0}+3\Phi_{3,1,2,1/3}+\Phi_{3,2,2,-2/3}+\Phi_{1,1,3,-2}$ & $(\frac{5}{3},\frac{4}{3},\frac{25}{6},\frac{29}{12})$ & $4 \times 10^{17}$ &  $10^{42 \pm 2.5}$\\ \hline                                                
\end{tabular}}
\end{center}
\caption{\label{tab:a.1}Simple LR configurations which can explain
  $[a.1]$ scalar CKM. One of the bi-doublets $\Phi_{1,2,2,0}$ is
  already considered in the basic field content
  (SM+bi-doublet). $m_{G}$ and $T_{1/2}$ have been calculated at
  1-loop. The first two configurations correspond to the minimal
  solutions, each one with the basic $[a.1]$ scalar CKM field
  content.}
\end{table}


\section{ SM-X extended unification: simple configurations}

It is possible to achieve one-step unification of the SM coupling
constants withing non-SUSY models. This is performed adding to the SM
a new particle content at scale $m_{NP}$. This particle content can be
as simple as the configurations shown in table~\ref{tab:X}, which
added to the $SM$ lead ``$SM+X$'' models that unify equal or even
better than the MSSM. Therefore, for each one of the fields in 
table~\ref{tab:List_of_LR_fields} (in the SM version) one of the simplest
$X$ configurations is obtained as follows:
$\alpha_{2}^{-1}(m_{G})-\alpha_{1}^{-1}(m_{G})<0.9$ (unification equal
or better that the MSSM) and $10^{15}<m_{G}<10^{18}$ GeV in order to
obtain proton life times allowed by the actual bounds.

\begin{table}[t]
\begin{center}
\scalebox{0.95}{
\begin{tabular}{| l | l | l |  l |}
\hline
Extra field & Configuration & $\Delta b^{'}s$ & $m_{G}$ \\ \hline

$\Phi_{1,2,1/2}$ & $\Phi_{1,2,1/2}+4\Phi_{3,2,1/6}+4\Phi_{3,1,1/3}$ & $(\frac{1}{2},\frac{13}{6},2)$ & $3 \times 10^{16}$ \\ 
               & $5\Phi_{1,2,1/2}+2\Phi_{1,3,0}+2\Phi_{8,1,0}$ & $(\frac{1}{2},\frac{13}{6},2)$ & $3 \times 10^{16}$  \\ \hline

$\Phi_{3,2,1/6}$ & $3\Phi_{3,2,1/6}$ & $(\frac{1}{10},\frac{3}{2},1)$ & $2 \times 10^{15}$  \\ 
               & $4\Phi_{3,2,1/6}+2\Phi_{3,1,-1/3}$ & $(\frac{4}{15},2,\frac{5}{3})$ & $8 \times 10^{15}$  \\ \hline

$\Phi_{3,1,2/3}$  & $4\Phi_{3,1,2/3}+2\Phi_{1,2,1/2}+5\Phi_{3,2,1/6}$ & $(\frac{43}{30},\frac{17}{6},\frac{17}{3})$ & $2 \times 10^{15}$ \\ 
               & $4\Phi_{3,1,2/3}+\Phi_{1,2,1/2}+5\Phi_{1,3,0}+3\Phi_{8,1,0}$ & $(\frac{7}{6},\frac{7}{2},\frac{11}{3})$ & $4 \times 10^{17}$  \\ \hline 
                                         
$\Phi_{3,1,-1/3}$ & $4\Phi_{3,1,-1/3}+\Phi_{1,2,1/2}+4\Phi_{3,2,1/6}$ & $(\frac{1}{2},\frac{13}{6},2)$ & $3 \times 10^{16}$ \\  
               & $4\Phi_{3,1,-1/3}+4\Phi_{1,2,1/2}+3\Phi_{1,3,0}+2\Phi_{8,1,0}$ & $(\frac{2}{3},\frac{8}{3},\frac{8}{3})$ & $1 \times 10^{17}$  \\ \hline
                             
$\Phi_{1,1,-1}$  & $3\Phi_{1,1,-1}+3\Phi_{1,2,1/2}+3\Phi_{1,3,0}+2\Phi_{8,1,0}$ & $(\frac{9}{10},\frac{5}{2},2)$ & $2 \times 10^{15}$  \\                
               & $\Phi_{1,1,-1}+2\Phi_{1,3,0}+2\Phi_{8,2,1/2}$ & $(\frac{9}{5},4,4)$ & $1 \times 10^{17}$  \\ \hline
                             
$\Phi_{3,1,0}$ & $3\Phi_{1,3,0}+2\Phi_{8,1,0}$ & $(0,2,2)$ & $1 \times 10^{17}$  \\ \hline
                                           
$\Phi_{8,1,0}$   &$2\Phi_{8,1,0}+3\Phi_{1,3,0}$ & $(0,2,2)$ & $1 \times 10^{17}$  \\ \hline
                           
$\Phi_{6,1,1/3}$ & $2\Phi_{6,1,1/3}+3\Phi_{1,3,0}$ & $(\frac{4}{15},2,\frac{5}{3})$ & $8 \times 10^{15}$ \\ \hline 
                      
$\Phi_{6,1,-2/3}$ & $2\Phi_{6,1,-2/3}+\Phi_{6,3,1/3}+\Phi_{3,2,-5/2}$ &  $(\frac{23}{10},\frac{9}{2},\frac{9}{2})$ & $1 \times 10^{17}$  \\  \hline
              
 $\Phi_{8,2,1/2}$ & $\Phi_{8,2,1/2}+3\Phi_{3,2,1/6}+\Phi_{1,2,1/2}$ &  $(1,3,3)$ & $1 \times 10^{17}$  \\ \hline
              
$\Phi_{1,3,-1}$ & $3\Phi_{1,3,-1}+3\Phi_{1,3,0}+4\Phi_{8,2,1/2}$ & $(\frac{9}{5},4,4)$ & $1 \times 10^{17}$ \\ \hline 
               
$\Phi_{1,1,-2}$ & $2\Phi_{1,1,-2}+2\Phi_{3,3,-1/3}+3\Phi_{8,1,0}$ &  $(2,4,4)$ & $1 \times 10^{17}$ \\ \hline
              
$\Phi_{3,2,7/6}$ & $\Phi_{3,2,7/6}+2\Phi_{1,3,0}+3\Phi_{8,2,1/2}+2\Phi_{1,3,-1}$ &  $(\frac{16}{3},\frac{22}{3},\frac{22}{3})$ & $1 \times 10^{17}$ \\ \hline
              
$\Phi_{3,3,-1/3}$ & $\Phi_{3,3,-1/3}+\Phi_{1,3,-1}+2\Phi_{8,1,0}$ & $(\frac{4}{5},\frac{8}{3},\frac{5}{2})$ & $3 \times 10^{16}$ \\ \hline
              
$\Phi_{3,1,-4/3}$ & $5\Phi_{3,1,-4/3}+2\Phi_{6,3,1/3}+2\Phi_{8,1,0}$ & $(\frac{92}{15},8,\frac{47}{6})$ & $3 \times 10^{16}$ \\ \hline
               
$\Phi_{6,3,1/3}$ & $\Phi_{6,3,1/3}+\Phi_{6,1,4/3}+\Phi_{8,2,1/2}$ & $(\frac{10}{3},\frac{16}{3},\frac{16}{3})$ & $1 \times 10^{17}$  \\ \hline
              
$\Phi_{6,1,4/3}$ & $\Phi_{6,1,4/3}+\Phi_{6,3,1/3}+\Phi_{8,2,1/2}$ & $(\frac{10}{3},\frac{16}{3},\frac{16}{3})$ & $1 \times 10^{17}$  \\ \hline
              
$\Phi_{3,2,-5/6}$& $\Phi_{3,2,-5/6}+4\Phi_{1,3,0}+3\Phi_{8,1,0}$ &  $(\frac{5}{6},\frac{19}{6},\frac{10}{3})$ & $4 \times 10^{17}$ \\ \hline
              
\end{tabular}}
\end{center}
\caption{\label{tab:X}Simple X configurations which lead ``SM+X''
  unification at $m_{G}$: $[10^{15},10^{18}]$ GeV. The first two
  configurations correspond to the examples described in 
  section~\ref{subsect:SlidLR}. Note that fields $\Phi_{3,1,-1/3}$,
  $\Phi_{3,1,-4/3}$, and $\Phi_{3,3,-1/3}$ are potentially dangerous
  for d=6 proton decay, see section~\ref{sec:requirements}.}
\end{table}


\begin{thebibliography}{10}


\bibitem{Dimopoulos:1981yj}
S.~Dimopoulos, S.~Raby, and F.~Wilczek,
\newblock Phys.Rev. {\bf D24}, 1681 (1981).

\bibitem{Ibanez:1981yh}
L.~E. Ibanez and G.~G. Ross,
\newblock Phys.Lett. {\bf B105}, 439 (1981).

\bibitem{Marciano:1981un}
W.~J. Marciano and G.~Senjanovic,
\newblock Phys.Rev. {\bf D25}, 3092 (1982).

\bibitem{Einhorn:1981sx}
M.~Einhorn and D.~Jones,
\newblock Nucl.Phys. {\bf B196}, 475 (1982).

\bibitem{Amaldi:1991cn}
U.~Amaldi, W.~de~Boer, and H.~Furstenau,
\newblock Phys.Lett. {\bf B260}, 447 (1991).

\bibitem{Langacker:1991an}
P.~Langacker and M.-x. Luo,
\newblock Phys.Rev. {\bf D44}, 817 (1991).

\bibitem{Ellis:1990wk}
J.~R. Ellis, S.~Kelley, and D.~V. Nanopoulos,
\newblock Phys.Lett. {\bf B260}, 131 (1991).

\bibitem{ArkaniHamed:2004fb}
N.~Arkani-Hamed and S.~Dimopoulos,
\newblock JHEP {\bf 0506}, 073 (2005), arXiv:hep-th/0405159.

\bibitem{Giudice:2004tc}
G.~F. Giudice and A.~Romanino,
\newblock Nucl.Phys. {\bf B699}, 65 (2004), arXiv:hep-ph/0406088.

\bibitem{Amaldi:1991zx}
U.~Amaldi, W.~de~Boer, P.~H. Frampton, H.~Furstenau, and J.~T. Liu,
\newblock Phys.Lett. {\bf B281}, 374 (1992).

\bibitem{Gogoladze:2010in}
I.~Gogoladze, B.~He, and Q.~Shafi,
\newblock Phys.Lett. {\bf B690}, 495 (2010), arXiv:1004.4217.

\bibitem{Brahmachari:1991np}
B.~Brahmachari, U.~Sarkar, and K.~Sridhar,
\newblock Phys.Lett. {\bf B297}, 105 (1992).

\bibitem{Mohapatra:1977mj}
R.~N. Mohapatra, F.~E. Paige, and D.~P. Sidhu,
\newblock Phys.Rev. {\bf D17}, 2462 (1978).

\bibitem{Mohapatra:1979ia}
R.~N. Mohapatra and G.~Senjanovic,
\newblock Phys.Rev.Lett. {\bf 44}, 912 (1980).

\bibitem{Mohapatra:1980yp}
R.~N. Mohapatra and G.~Senjanovic,
\newblock Phys.Rev. {\bf D23}, 165 (1981).

\bibitem{Mohapatra:1986bd}
R.~N. Mohapatra and J.~W.~F. Valle,
\newblock Phys. Rev. {\bf D34}, 1642 (1986).

\bibitem{Akhmedov:1995ip}
E.~K. Akhmedov, M.~Lindner, E.~Schnapka, and J.~W.~F. Valle,
\newblock Phys.Lett. {\bf B368}, 270 (1996), arXiv:hep-ph/9507275.

\bibitem{Akhmedov:1995vm}
E.~K. Akhmedov, M.~Lindner, E.~Schnapka, and J.~W.~F. Valle,
\newblock Phys. Rev. {\bf D53}, 2752 (1996), hep-ph/9509255.

\bibitem{Brahmachari:2003wv}
B.~Brahmachari, E.~Ma, and U.~Sarkar,
\newblock Phys.Rev.Lett. {\bf 91}, 011801 (2003), arXiv:hep-ph/0301041.

\bibitem{Siringo:2012bc}
F.~Siringo,
\newblock Phys.Part.Nucl.Lett. {\bf 10}, 94 (2013), arXiv:1208.3599.

\bibitem{Aulakh:1997ba}
C.~S. Aulakh, K.~Benakli, and G.~Senjanovic,
\newblock Phys.Rev.Lett. {\bf 79}, 2188 (1997), arXiv:hep-ph/9703434.

\bibitem{Aulakh:1997fq}
C.~S. Aulakh, A.~Melfo, A.~Rasin, and G.~Senjanovic,
\newblock Phys.Rev. {\bf D58}, 115007 (1998), arXiv:hep-ph/9712551.

\bibitem{Esteves:2011gk}
J.~N. Esteves {\em et~al.},
\newblock JHEP {\bf 1201}, 095 (2012), arXiv:1109.6478.

\bibitem{Malinsky:2005bi}
M.~Malinsky, J.~C. Romao, and J.~W.~F. Valle,
\newblock Phys.Rev.Lett. {\bf 95}, 161801 (2005), arXiv:hep-ph/0506296.

\bibitem{Majee:2007uv}
S.~K. Majee, M.~K. Parida, A.~Raychaudhuri, and U.~Sarkar,
\newblock Phys.Rev. {\bf D75}, 075003 (2007), arXiv:hep-ph/0701109.

\bibitem{Dev:2009aw}
P.~S.~B. Dev and R.~N. Mohapatra,
\newblock Phys.Rev. {\bf D81}, 013001 (2010), arXiv:0910.3924.

\bibitem{DeRomeri:2011ie}
V.~De~Romeri, M.~Hirsch, and M.~Malinsky,
\newblock Phys.Rev. {\bf D84}, 053012 (2011), arXiv:1107.3412.

\bibitem{Arbelaez:2013hr}
C.~Arbelaez, R.~M. Fonseca, M.~Hirsch, and J.~C. Romao,
\newblock Phys.Rev. {\bf D87}, 075010 (2013), arXiv:1301.6085.

\bibitem{Lindner:1996tf}
M.~Lindner and M.~Weiser,
\newblock Phys.Lett. {\bf B383}, 405 (1996), arXiv:hep-ph/9605353.

\bibitem{Bertolini:2009es}
S.~Bertolini, L.~Di~Luzio, and M.~Malinsky,
\newblock Phys. Rev. {\bf D81}, 035015 (2010), arXiv:0912.1796 [hep-ph].

\bibitem{Bertolini:2013vta}
S.~Bertolini, L.~Di~Luzio, and M.~Malinsky,
\newblock Phys.Rev. {\bf D87}, 085020 (2013), arXiv:1302.3401 [hep-ph].

\bibitem{Calmet:2008df}
X.~Calmet, S.~D. Hsu, and D.~Reeb,
\newblock Phys.Rev.Lett. {\bf 101}, 171802 (2008), arXiv:0805.0145 [hep-ph].

\bibitem{Dvali:2007hz}
G.~Dvali,
\newblock Fortsch.Phys. {\bf 58}, 528 (2010), arXiv:0706.2050 [hep-th].

\bibitem{Babu:2012vb}
K.~S. Babu and R.~N. Mohapatra,
\newblock Phys.Rev. {\bf D86}, 035018 (2012), arXiv:arXiv:1203.5544.

\bibitem{Weinberg:1980bf}
S.~Weinberg,
\newblock Phys. Rev. {\bf D22}, 1694 (1980).

\bibitem{Weldon:1980gi}
H.~A. Weldon and A.~Zee,
\newblock Nucl.Phys. {\bf B173}, 269 (1980).

\bibitem{Abe:2013lua}
Super-Kamiokande Collaboration, K.~Abe {\em et~al.},
\newblock (2013), arXiv:1305.4391.

\bibitem{Buras:1977yy}
A.~J. Buras, J.~R. Ellis, M.~K. Gaillard, and D.~V. Nanopoulos,
\newblock Nucl.Phys. {\bf B135}, 66 (1978).

\bibitem{Ellis:1979hy}
J.~R. Ellis, M.~K. Gaillard, and D.~V. Nanopoulos,
\newblock Phys.Lett. {\bf B88}, 320 (1979).

\bibitem{Wilczek:1979hc}
F.~Wilczek and A.~Zee,
\newblock Phys.Rev.Lett. {\bf 43}, 1571 (1979).

\bibitem{Nishino:2012ipa}
Super-Kamiokande, H.~Nishino {\em et~al.},
\newblock Phys.Rev. {\bf D85}, 112001 (2012), arXiv:1203.4030.

\bibitem{ATLAS:2013oma}
ATLAS Collaboration,
\newblock (2013).

\bibitem{Chatrchyan:2013lba}
CMS Collaboration, S.~Chatrchyan {\em et~al.},
\newblock JHEP {\bf 1306}, 081 (2013), arXiv:1303.4571.

\bibitem{Beringer:1900zz}
Particle Data Group, J.~Beringer {\em et~al.},
\newblock Phys.Rev. {\bf D86}, 010001 (2012).

\bibitem{Amsler:2008zzb}
Particle Data Group, C.~Amsler {\em et~al.},
\newblock Phys.Lett. {\bf B667}, 1 (2008).

\bibitem{Jones:1981we}
D.~R.~T. Jones,
\newblock Phys.Rev. {\bf D25}, 581 (1982).

\bibitem{Machacek:1983tz}
M.~E. Machacek and M.~T. Vaughn,
\newblock Nucl.Phys. {\bf B222}, 83 (1983).

\bibitem{Luo:2002ti}
M.-x. Luo, H.-w. Wang, and Y.~Xiao,
\newblock Phys.Rev. {\bf D67}, 065019 (2003), arXiv:hep-ph/0211440.

\end{thebibliography}
\end{document}